# Ballistic Energy Transport via Long Alkyl Chains: A New Initiation Mechanism

Sithara U. Nawagamuwage, Elliot S. Williams, Md Muhaiminul Islam, Igor V. Parshin, Alexander L. Burin, Nathalie Busschaert, Igor V. Rubtsov*

*Department of Chemistry, Tulane University, New Orleans, Louisiana 70118, United States*

irubtsov@tulane.edu

**Abstract**

In an effort to increase the speed and efficiency of ballistic energy transport via oligomeric chains, we performed measurements of the transport in compounds featuring long alkyl chains of up to 37 methylene units. Compounds of the $N_3$-$(CH_2)_n$-COOMe type (denoted as az*n*ME) were synthesized with *n* = 5, 10, 15, 19, 28, 37 and studied using relaxation-assisted two-dimensional infrared spectroscopy. The speed of the ballistic transport, initiated by the $N_3$ tag excitation, increased ca. 3-fold for the longer chains (*n* = 19-37) compared to the shorter chains, from 14.7 Å/ps to 48 Å/ps, in line with an earlier prediction (Nawagamuwage et al. 2021, *J. Phys. Chem. B*, **125**, 7546). Modeling, based on solving numerically the Liouville equation, was capable of reproducing the experimental data only if three wavepackets are included, involving $CH_2$ twisting (Tw), wagging (W), and rocking (Ro) chain bands. The approaches for designing molecular systems featuring higher speed and efficiency of energy transport are discussed.

## 1. Introduction

Vibrational energy flow in molecules has been a topic of interest in chemistry since Arrhenius's explanation of heat dependence of reaction rates.[1, 2] Even though in nature the most common form of heat transfer involves diffusive hops of vibrational energy by Brownian motion, there is the possibility to transfer vibrational energy ballistically in condensed matter. This requires strong coupling of a number of local states extended over significant range of distances forming delocalized states.[3-11] Particularly it was shown recently that vibrational energy delivery controls enzyme functioning in soybean lipoxygenase[12] and intersystem crossing rate in iron(II)-tris(bipyridine).[13] The presence of strong covalent bonds in periodic molecular chains results in generally strong interactions of nearest neighbor local sites, facilitating long-range delocalization. Such delocalized coupled states are traditionally described by dispersion relations, $\omega(q)$, where $\omega$ is the frequency (energy) of the coupled states and $q$ is its wavevector.[8, 14] Acoustic phonons in materials and molecular chains are responsible for heat conduction.[15, 16] Optical phonons can be used to transfer larger quanta of energy, released e. g. in chemical reactions, ballistically to longer distances, making such systems potentially useful as new materials for heat management, molecular electronics and sensors.[17-20] Recent work[21, 22] suggests similar applications of vibrational polaritons which are quantum superpositions of vibration and infrared light.

Dispersion curves for infinitely long linear alkyl chains involving $CH_2$ scissoring, wagging (W), twisting (Tw), and rocking (Ro) bands and C-C stretching band are shown in Figure 1. The widths of different bands and the shapes of the dispersion curves differ as interaction strengths and signs of the local vibrational states of different types vary. Actual energies of coupled states of a band for a chain of finite length can be obtained from its dispersion curve. By exciting a superposition of two or more delocalized chain states a vibrational wavepacket is formed.

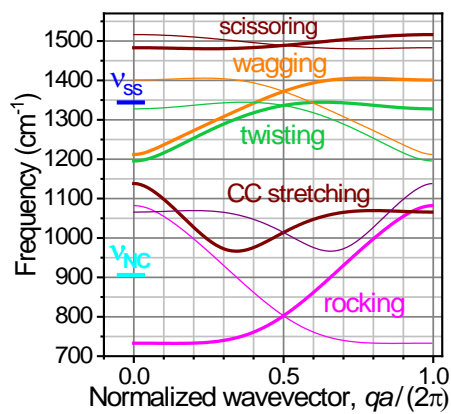

**Figure 1.** Dispersion curves for five optical bands of linear alkyl chains. Vibrational frequencies of the azido moiety of the $N_3$-alkyl compound, the $N_3$ asymmetric stretch ($\nu_{ss}$) and $\nu_{NC}$, are also shown.[23]

Once formed, the wavepacket propagates in the chain with time, moving its energy along the chain at a nearly constant speed. Such energy transport is called ballistic transport; it is typically much faster and more efficient than diffusive transport.[14] The transport speed of the wavepacket formed in the vicinity of the wavevector $q_0$ (its group velocity at $q_0$) is determined by the slope of the dispersion curve: $V(q_0) = (\partial\omega/\partial q)|_{q=q_0}$. Depending on how wide the range of the wavevectors involved in the wavepacket is, the wavepacket transport speed can vary significantly. Therefore, understanding the process of formation of the wavepacket in the chain is of ultimate importance for describing the energy transport in such chains.

Ballistic energy transport via covalent bonds in molecules was studied actively over the past two decades. The transport was initiated by electronic excitation (~2 eV)[24, 25] or by vibrational excitation of an end group of oligomeric chains (1500-2100 cm$^{-1}$).[26] Various chain types were investigated following vibrational excitation and detection using relaxation-assisted two-dimensional infrared (RA 2DIR) spectroscopy.[27, 28] In these experiments, the energy transport was initiated by exciting a high-frequency mode of a chain end-group (the tag). Vibrational relaxation of the excited mode initiates a wavepacket, which propagates freely in the chain. The arrival of the wavepacket to another end of the chain is recorded via 2DIR as a change in the frequency of the reporter mode located at this end of the chain caused by the excess energy arrival. The transport speed is measured as an inverse slope of the dependence of the energy arrival time as a function of the chain length.

The transport speed was found to be dependent on the chain type (PEG,[26, 29-31] perfluoroalkyl,[32] alkyl,[23, 33, 34] and oligo-*p*-phenylene[35]) and on the selection of the tag mode initiating the transport but not on the reporter mode selection. For example the speed of 80 Å/ps (8 km/s) was observed in oligophenylene chains,[35] which is ca. 5-times larger than the speed of sound in water. The transport speed via linear alkyl chains up to 15 methylene units, measured using RA 2DIR, was found to be strongly dependent on the type of the end-group initiating the transport with 14.4, 8.0, and 4.3 Å/ps for the $N_3$ (2100 cm$^{-1}$), NHS-ester (1740 cm$^{-1}$), and amide (1640 cm$^{-1}$) mode initiations, respectively, associated with transport via different optical bands of the chain.[33]

A recent comprehensive study on the wavepacket initiation mechanisms using azido groups in alkyl chains of length up to 15 repeating units identified several mechanisms of wavepacket initiation.[36] It was shown that the asymmetric N≡N stretching mode ($\nu_{as}$, 2100 cm$^{-1}$, the tag), relaxes predominantly into a combination band of the symmetric $N_3$ stretching mode, $\nu_{ss}$, and N-C stretching mode, $\nu_{NC}$ (Fig. 1, 2).[23, 37] Fast intramolecular vibrational energy redistribution (IVR) process of $\nu_{ss}$ populates a range of Tw chain states as well a low frequency mode, creating a wavepacket in the Tw band (Fig. 2A, 2C left), which is the dominant mechanism of wavepacket formation for short chains. It was predicted that for longer chains ($n > 15$), additional mechanisms of wavepacket formation can become efficient resulting in wavepackets in the W and Ro bands. The new wavepacket formation channels are expected to become efficient at increased chain-state density so that more than one W chain state can be found within $v_{ss}$ ($n > 15$) and more than one Ro state is found within $v_{CN}$ ($n > 25$), Figure 2C, right.

Here we report on RA 2DIR studies (Section 3.1) and density matrix modeling (Section 3.2) of the energy transport in molecular wires of $^{14}N_3(CH_2)_n$-COOMe type (denoted as az$n$ME) with $n$ = 5, 10, 15, 19, 28, and 37. The discussion section (Section 4) describes the possible steps to increase the speed, efficiency of initiation, and efficiency of ballistic transport.

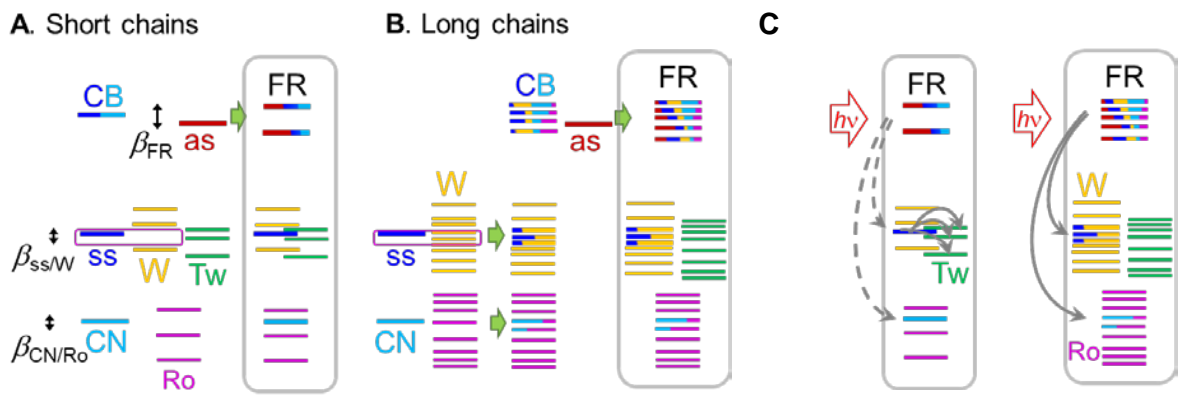

**Figure 2. A-B**. Energy schematics showing harmonic mixing of the end-group and chain states and nonlinear mixing of $v_{as}$ with the combination band (CB) of predominantly $v_{ss}$ and $v_{CN}$ for the cases of short (**A**) and long (**B**) chains. The interaction energies, $\beta_{ss/W} \sim 10$ cm$^{-1}$, $\beta_{CN/Ro} \sim 9$ cm$^{-1}$, $\beta_{FR} \sim 30$ cm$^{-1}$, are indicated with double-sided arrows ($\beta_{ss/Tw} < 0.1$ cm$^{-1}$). Formation of the harmonically mixed end-group and chain states is shown in the single-quantum manifold within grey boxes. The resulted combination bands involving $v_{ss}$ and $v_{CN}$ local modes are labeled as CB. The nonlinear mixture of the CB with $v_{as}$ results in the coupled two quanta states denoted as FR. The IR intensities of these states (GS + $h\nu \to$ FR) scale as the square of the contribution of $v_{ss}$ in the coupled state wavefunction. **C.** Mechanisms of wavepacket formation for short (left) and long (right) chains.

## 2. Experimental Details

***Synthesis.*** A series of compounds featuring linear alkyl chains of lengths $n$ = 5, 10, 15, 19, 28, and 37 each terminated with one azido and one methyl ester end groups, denoted as az$n$ME, were synthesized. For $n$ = 5, 10 and 15, the compounds could be easily obtained using a substitution reaction from commercially available bromo-terminated carboxylic acids with NaN$_3$, followed by esterification to methyl esters. The longer chains ($n$ = 19, 28 and 37) had to be synthesized in a stepwise fashion, using alkene cross-metathesis to elongate the chains. Thus, az19ME was obtained from the cross-metathesis of a C11 carboxylate ester containing a terminal alkene and 11-bromoundecene, followed by a reduction of the internal alkene and substitution of the terminal bromine with azide. Similarly, az28ME was obtained from the cross-metathesis of one of the az19ME intermediates with 11-bromoundecene, followed by a reduction of the internal alkene and substitution of the terminal bromine with azide; while az37ME was obtained from the cross-metathesis of one of the az28ME intermediates with 11-bromoundecene, followed by a reduction of the internal alkene and substitution of the terminal bromine with azide. A detailed reaction scheme, reaction conditions and characterization are given in the SI.

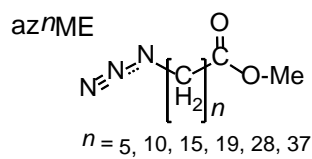

**Scheme 1.** Structure of the az*n*ME compounds.

***Sample preparation.*** For FTIR and 2DIR measurements, each compound was dissolved in $CDCl_3$ to prepare a 30 mM solution. All experiments were performed in a sample cell with a 100 µm Teflon spacer separating two 1-mm-thick $CaF_2$ windows.

***2DIR measurements.*** We used relaxation-assisted two-dimensional infrared (RA 2DIR) spectroscopy[38] to study energy transport along alkyl chains. A comprehensive description of the fully automated dual frequency three-pulse photon echo 2DIR spectrometer used is formerly published elsewhere[39]. Briefly, an output of 800 nm wavelength from a Ti:Sapphire laser producing 1.5 W power at 1 kHz repetition rate with 80 fs pulse duration (Libra, Coherent) was passed through a dual optical parametric amplifier (OPA, Palitra-duo, Quantronix) and a pair of difference frequency generation (DFG) units (NIR Quantronix) to generate tunable mid-IR pulses in the frequency range from 500 to 5000 $cm^{-1}$, featuring pulse energy ranging from 1.0 to 10 µJ. The pulse energies of the mid-IR beams at the sample were 1.5 and 1 µJ at 2100 and 1730 $cm^{-1}$, respectively.

The instrument is equipped with a mid-IR beam direction stabilization schematic[40] and a schematic for setting the phase-matching geometry for mid-IR beams at the sample[39] to enable automated navigation to any cross or diagonal 2DIR peak. The spectral width of the mid-IR pulses was ~150 $cm^{-1}$, and the instrument response function was ~180 fs.

The 2DIR measurements were performed by scanning the delay between the first two mid-IR pump pulses, ($\tau$) originated from the same DFG unit, at a fixed waiting time ($T$), which is the delay between the second and third pulses, and recording the heterodyned spectrum in the frequency range of interest ($\lambda \rightarrow \omega_t$) for every $T$. Fourier transformation along $\tau$ results in the $\omega_\tau$ axis in the 2DIR spectrum, shown as the ordinate. A typical 2DIR spectrum contained ~250 points along the $\tau$ direction, which took 1–3 min to acquire. For RA 2DIR measurements, the waiting time, $T$, was scanned with nonconstant delay steps ranging from 100 fs at small waiting times up to 5 ps at large waiting times. Typical waiting-time dependences contained 40–50 points along $T$, which took ca. 1.5 hours to acquire.

***Theoretical modeling.***

We described the transport by solving Liouville - Bloch equations for the density matrix as in Ref. [41]. The hopping through the chain of *n* sites, enumerated by numbers 1, 2, ..., *n*, was characterized by nearest neighbor and next neighbor couplings, $\beta$ and $\gamma$, respectively, as introduced below in Eq. 1, and dephasing and relaxation rates, $k_{deph}$ and $k_{rlx}$. The first site of the chain is coupled to the initiation zeroth site, characterized by the energy $E_0$ with the coupling strength $\beta_{tag}$. The last site, *n*, is coupled to the reporter, whose population obeys the kinetic equation $dP/dt = k_{rep} \rho_{nn} - P / T_c$, where $P$ is the population of the reporter, $k_{rep}$ is the rate of the energy transfer from the $n^{th}$ site of the chain to the reporter state, $\rho_{nn}$ is the diagonal element of the density matrix, characterizing the excitation density at the site adjacent to the reporter and $T_c$ is the reporter cooling rate to the solvent. We do not assume that the reporter C=O stretching mode is populated because its frequency exceeds those of the optical bands responsible for the energy transport. There are several lower frequency modes at the reporter end-group coupled anharmonically to the reporter mode. Excitation of those modes shift the reporter mode energy. We assume that the cross-peak intensity is defined by the population of the reporter $P(t)$; the energy transport time was evaluated as time where the $P(t)$ trace reaches its maximum. The initial conditions were set similarly to that in Ref. [41] as $\rho_{ab}(0) = \delta_{a0} \delta_{b0}$, where $\delta_{ab}$ is a Kronecker symbol.

For the transport involving several chain bands, contributions from individual bands were added with appropriate weight factors, as described in Section 3. We consider harmonic initiation of the wavepacket. In the case of its initiation by means of anharmonic transition the wavepacket is formed if the initiation emerges at a sufficiently short time scale,[42] as shown for wavepacket initiation in the Tw band.[36]

The present consideration is still relevant even for the anharmonic initiation since the most important observations are associated with the wavepacket transport, rather than its initiation.

## 3. Results

### 3.1. RA 2DIR measurements of the energy transport in alkyl chains, initiated by $\nu_{N\equiv N}$.

In the set of experiments, the energy transport was initiated by exciting asymmetric stretch of the azido group at 2100 cm$^{-1}$ ($\nu_{N\equiv N}$, the tag) at one end of the molecule (Fig. 3A). The excess energy, introduced to the azido group, relaxes to the alkyl chain, resulting in wavepacket formation. The wavepacket propagates through the chain and arrives at the other end of the molecule, wherein it is observed by the carbonyl stretch of the methyl ester group ($\nu_{C=O}$, the reporter) at 1730 cm$^{-1}$ (Fig. 3A).

The energy arrival to the methyl ester moiety is observed as an amplitude increase of the cross peak between the tag and reporter (Fig. 3B). The amplitude of the $\nu_{N\equiv N}$ / $\nu_{C=O}$ cross-peak in the 2DIR spectrum increases with the waiting time due to the arrival of excess energy to the reporter site (Fig. 4). For each waiting time, the $\nu_{N\equiv N}$ / $\nu_{C=O}$ cross-peak area was integrated and plotted as a function of the waiting time. The resulting waiting time traces are shown in Figure 4 for three indicated compounds. The traces were fitted with an asymmetric double–sigmoidal function (Fig. S31 caption) and the curve maximum was determined from the fit. The time taken to reach the highest cross-peak amplitude is referred to as an energy transport time, $T_{max}$. Note that no less than three experiments were performed for each sample (Fig. 4) and the reported $T_{max}$ value for each sample were obtained by averaging those results. The averaged $T_{max}$ values were plotted as a function of the tag-reporter distance (Figure 5). Finally, the energy transport speed can be determined as the inverse slope of the $T_{max}(R)$ dependence.

As apparent from Figure 5, the $T_{max}(R)$ dependence is not linear over the whole range of chain lengths but shows a clear kink at $R \sim 32$ Å. The experimental speed obtained from the linear fit of the data for short chains ($n < 19$) is found at $14.7 \pm 0.3$ Å/ps, which accords with the previous studies with different reporter groups, such as NHS ester ($14.7 \pm 0.3$ Å/ps),[23] carboxylic acid ($14.4 \pm 2$ Å/ps),[36] and an amide ($14.0 \pm 0.6$ Å/ps for amide-I and $13.9 \pm 0.6$ Å/ps for amide-II).[34] While the speeds recorded with different reporters were the same, the $T_{max}$ values for the chains of the same lengths were different for different reporters, so the whole series in this study was performed with the same reporter – methyl ester. A linear fit of the points for the longer chains ($n \geq 19$), resulted in the transport speed of $48 \pm 3$ Å/ps which is ca. 3 times faster than that for the short chains, suggesting that the mechanism of the transport changes for longer chains.

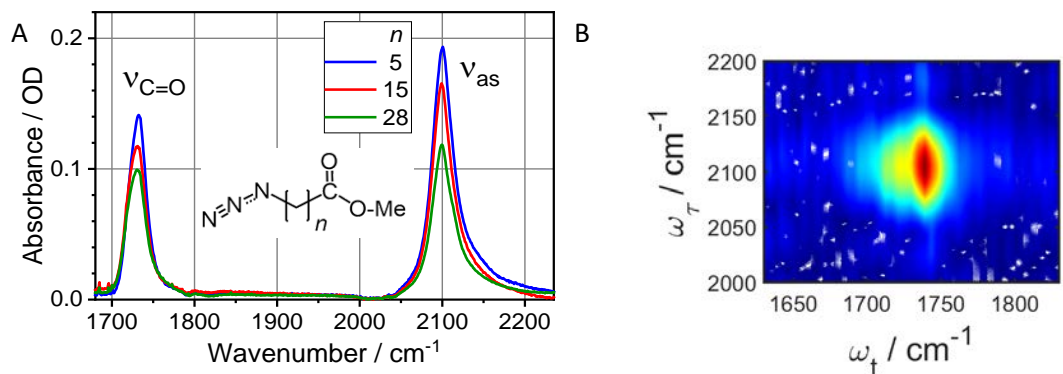

**Figure 3. A.** Linear spectra of az$n$ME ($n$ = 5, 15, 28) compounds in CDCl$_3$. $\nu_{as}$ of the N$_3$ moiety and $\nu_{C=O}$ of the methyl ester moiety are indicated. **B.** 2DIR rephasing spectrum of az28Me at $T$ = 5.5 ps focusing at the $\nu_{N\equiv N}/\nu_{C=O}$ cross peak.

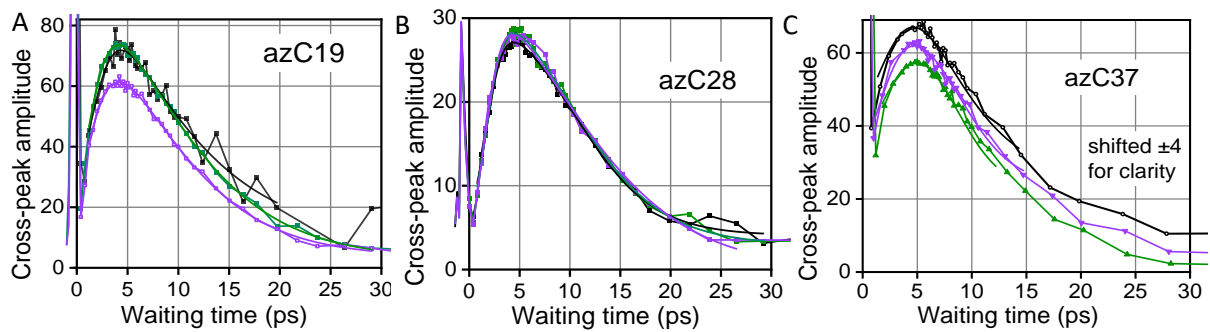

**Figure 4.** Waiting time traces for the $\nu_{N\equiv N}/\nu_{C=O}$ cross-peak for (A) az19ME, (B) az28ME, and (C) az37ME. Three traces are shown for each compound. The fit curves with an asymmetric double–sigmoidal function (see Fig. S31 in SI), are shown with matching colors.

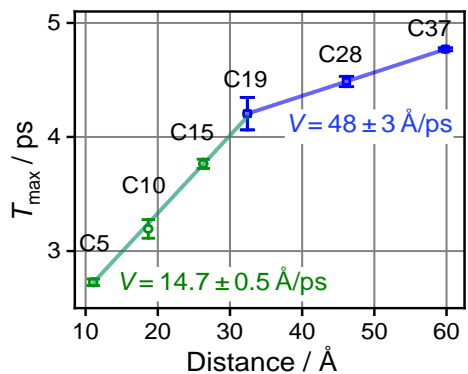

**Figure 5.** Dependence of $T_{max}$ on the tag reporter distance. Linear fits for the az5ME – az19ME and az19ME – az37ME are shown; the resulted energy transport speeds are shown as inset with matching colors.

### 3.2. Modeling of the wavepacket transport

A model based on solving the Liouville - von Newman equation for the density matrix was developed, coded in MATLAB, and numerically solved. The az$n$ME chain was represented by $n$ local states interacting with each other via nearest, $\beta$, and next nearest, $\gamma$, neighbor interaction coupling constants. The tag end group was represented by a single state coupled to the nearest chain state, 1, with $\beta_{tag}$. Numerical solution of the density matrix equation results in the time evolution of the diagonal density matrix elements, which represent excess energy at different sites. The reporter site is added in the ad-hoc fashion as a relaxation channel from the last site of the chain with a rate constant, $k_{report}$.

The parameters of the chain band were taken from the fit of the computed chain states for infinitely long chain data with an analytical function given by Equation 1.[36, 43]

$$\omega(q) = \omega_0 + 2\beta \cos(qa) + 2\gamma \cos(2qa) \tag{1}$$

Here $\omega_0$ is the site energy, $a$ is the chain period, and $q$ is the wavevector of the associated delocalized normal mode.

Previous studies suggested that for the compounds with short chains ($n \leq 15$), the wavepacket at the CH$_2$ twisting (Tw) band dominated the transport.[23] However, for longer chains, formation of the wavepacket at CH$_2$ wagging (W) band is expected, which would result in an increased speed.[36] Furthermore, a wavepacket at the CH$_2$ rocking band (Ro) may contribute for even longer chains ($n > 30$), characterized by a significantly faster speed.

**Table 1.** Chain band parameters for twisting, wagging, and rocking bands: site energy ($E_0$) and nearest ($\beta$) and next-nearest ($\gamma$) coupling constants.[23, 36]

| Chain bands | $\omega_0$, cm$^{-1}$ | $\beta$, cm$^{-1}$ & | $\gamma$, cm$^{-1}$ & | Maximal group velocity, Å/ps |
|---|---|---|---|---|
| Wagging | 1344 | -43 | -15 | 35 |
| Twisting | 1303 | -28 | -18 | 26 |
| Rocking | 852 | -83.8 | 25.7 | 69 |

&The error bars for $\beta$ and $\gamma$, originated from the fit of the dispersion curves with Eq. 1, were not exceeding 0.5 cm$^{-1}$. However, the computational DFT errors for the dispersion curves are expected to be larger, within ±3 cm$^{-1}$. The error bars for $E_0$ with respect to the end-group state energies $\nu_{ss,N3}$ and $\nu_{C-N}$ were estimated to be within 10 cm$^{-1}$.

**3.3. Transport via a single chain band.** We first investigated if the modeling involving a single chain band, Tw, can reproduce the experimental data. We found that while the kink in the $T_{max}(R)$ dependence can be obtained in the calculations, the computed transport speed for the longer chains, limited by the maximum speed supported by the Tw band, is significantly smaller than the experimental speed. The $\beta$ and $\gamma$ parameters determine the width of the chain band and thus the maximum group velocity supported by the band (Table 1). We kept $\beta$ and $\gamma$ fixed with the values determined from the fit with Eq. 1 of previously computed dispersion curve for the Tw band (see Table 1).[36]

To clearly observe the wavepacket motion, we first set the modeling parameters so that the wavepacket propagation in the chain occurs with minimal losses inside the chain as well as small losses to the reporter site. To achieve that, we set the rates of dephasing ($k_{deph}$), relaxation ($k_{rlx}$), and leak to the reporter ($k_{rep}$) to be negligible. The cooling process of the reporter site was eliminated as well (large $T_c$). Under these conditions, the energy at the reporter site, $S(T)$, accumulates slowly in a step like fashion (Fig. 6, red line). To emphasize the steps in $S(T)$, the derivative of $S(T)$, $\partial S/\partial T$, is shown in Figure 6 with blue line. The peaks in $\partial S/\partial T$ match the energy leakage bursts from the chain to the reporter site. Notice that the wavepacket undergoes some broadening and deformations due to nonlinearity of the dispersion relations for optical bands. The repeating peaks feature time separation corresponding to the wavepacket round trip in the chain. Despite the band-state nonlinearity, it is apparent that the wavepacket can potentially exist in the chain for a long time, exceeding 10 ps, given that the dephasing and relaxation of the chain states is slow. It is also clear that the peak in the $S(T)$ dependence (Fig. 4), is formed via a competition of the energy arrival to the reporter site and the reporter site cooling.[44]

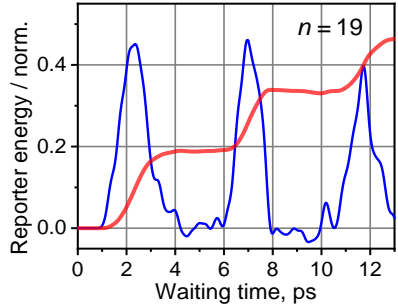

**Figure 6**. Computed excess energy at the reporter site (($S(T)$, red line) transported via the Tw band of the chain with $n$ = 19 and its numerical time derivative (blue line) as a function of time when $k_{deph}$ = 0 ps$^{-1}$, $k_{rlx}$ = 0 ps$^{-1}$, $k_{rep}$ = 0.05 ps$^{-1}$, $E_0$ = -25 cm$^{-1}$, and $\beta_{tag}$ = 10 cm$^{-1}$.

By using realistic parameters for the wavepacket motion, a good match of the modeling and the experiment was achieved but only for short chains ($n$ = 3-19), Figure 7. The modeling parameters used are shown in Figure 7 caption. The modeling describes well the presence and the location and the kink in the $T_{max}(R)$ dependence, while the speed for longer chains is not well described.

The $T_{max}$ value of the kink in the $T_{max}(R)$ dependence of ca. 3.5 ps dictates the selection of the values for $k_{deph}$ and $k_{rlx}$ of about (2-3 ps)$^{-1}$. Under these conditions, additional energy roundtrips in the chain, caused by the wavepacket reflection at the chain end and then second reflection at the tag site, became inefficient due to dephasing. Even for a short chain of $n = 5$, the second wavepacket arrival after a roundtrip occurs at ca. 4 ps, delivering twenty-fold smaller energy compared to delivered at the first wavepacket arrival, still affecting the waiting-time dependence and the $T_{max}$ value. The round trip for longer chains takes longer, making the amount of energy delivered after the wavepacket round trip even smaller. As a result, an increase in the apparent transport speed can be observed even for a single band. However, the maximum apparent, $T_{max}$-based speed is still limited by the maximum speed supported by the band.

The energy of the tag state with respect to the chain band, $E_0$, affects minimally the $T_{max}(R)$ values for all chain lengths. Note that the DFT computations place the $\nu_{ss,N3}$ energy close to the top of the Tw band (Fig. 1). However, the transport speed in short chains is affected by $\beta_{tag}$, especially when the tag energy, $E_0$, is close to the upper portion of the chain band. A small density of chain states in the short chains results in involvement of only a few chain states in the wavepacket. $\beta_{tag}$ determines the number and identity of these states, affecting the transport speed: larger $\beta_{tag}$ values make the apparent speed faster (reaching 18 Å/ps). Reducing $\beta_{tag}$ below 5 cm$^{-1}$ results in a lower apparent speed as well as in larger $T_{max}$ values, which exceed the experimental values significantly (~4 ps already for $n = 5$). Small $k_{rep}$ values (< 1 ps$^{-1}$) also lead to excessive $T_{max}$ values, restricting $k_{rep}$ to the range of 2 - 4 ps$^{-1}$.

The parameters resulting in a good match with the experimental data at smaller $n$ and reproducing the kink in the dependence, obtained using guess and check method, are shown in the Figure 7 caption. The magenta line in Figure 7, obtained as $T_{max}$ of $\partial S/\partial T$, represents the delay at which the energy flux into the reporter is maximal. The wavepacket speed determined from it, 26.4 Å/ps, represents the maximal speed supported by the band, which is ca. 26 Å/ps. Importantly, the dependence is essentially linear over the whole range of chain lengths, indicating that even if $\beta_{tag}$ is smaller than the intrachain couplings ($\beta$ and $\gamma$), multiple chain states participate in the wavepacket. While the modeling parameters obtained are not unique, the range of their acceptable values is not very wide.

For longer chains, the speed always gravitates to $V_{max}$ staying within the 24-26 Å/ps range. No conditions were found in the modeling to obtain a faster speed than the maximal speed supported by the Tw band. Therefore, we conclude that the energy transport via Tw band alone cannot describe the experiment for all chain lengths. We next consider the transport involving the Tw and W bands together and the transport involving Tw, W, and Ro bands.

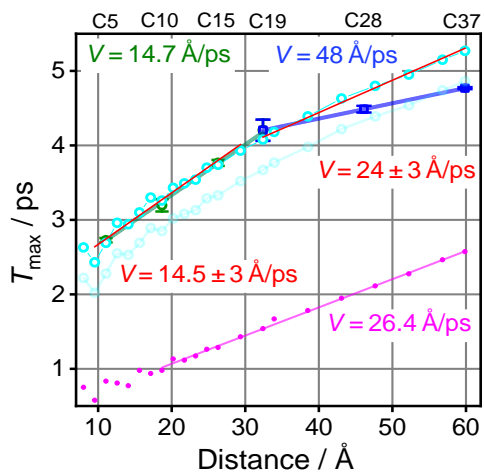

**Figure 7.** (left) Dependence of $T_{max}$ on the tag reporter distance, duplicated from Figure 5 (green and blue symbols and lines), overlayed with the modeling (cyan circles) for different chain lengths. The light cyan circles are the computed values; the data shown with cyan color are shifted up by 0.45 ps, accounting for the time of a wavepacket formation. Linear fits of the modeling data for the $n$ of 3 – 19 and $n$ of 19-37 are shown with thin red lines and respective speeds are indicated as insets in red. The modeling parameters were: $E_0$ = -25 cm$^{-1}$, $\beta_{tag}$ = 10 cm$^{-1}$, $k_{deph}$ = 0.27 ps$^{-1}$, $k_{rlx}$ = 0.5

ps$^{-1}$, $k_{rep}$ = 4 ps$^{-1}$, and $T_c$ = 10 ps. The maximum energy flow to the reporter, determined as the maximum of $\partial S/\partial T$, is shown with magenta dots and its linear fit with magenta line. (right) Experimental waiting time dependence for az19ME (blue) and modelled dependence for $n$ = 19 (green). To reduce the fluctuations in the $T_{max}$ values, we incorporated inhomogeneity for the energies of the tag state of 10 cm$^{-1}$ (see Section S2 in SI for more details).

### 3.4. Transport involving two or three chain bands.

It was previously reported that the W modes of the chain are harmonically coupled to $\nu_{ss}$ with the interaction energy of ca. 10 cm$^{-1}$.[36] With the large width of the W band and because the $\nu_{ss}$ energy falls into the region of the highest slope of the W band, only a small mixing of $\nu_{ss}$ and W states occurs for short chains with $n$ < 15, making the wavepacket on the W band insignificant for such chains. The Ro band is significantly broader than the W band and $\nu_{CN}$ is located at the very steep portion of the Ro band (Fig. 1) but their coupling is only ca. 9 cm$^{-1}$, so the wavepacket within the Ro band is expected to contribute at even larger chain lengths.

The calculations were performed with two or three independent wavepackets and summing the energy delivered to the reporter site by each wavepacket. Each band was represented by its $\beta$ and $\gamma$ values (Table 1), interaction to the tab ($\beta_{tag}$), dephasing ($k_{deph}$) and relaxation ($k_{rlx}$) rates, and the rate of populating the reporter site ($k_{rep}$). In addition, contributions of each wavepacket, involving the relative efficiencies of the wavepacket formation, were used as fit parameters. Figure 8 shows the results of modeling by using two chain bands, Tw and W. The kink in the $T_{max}(R)$ dependence occurs when the transport via W band becomes more efficient than that via Tw band. As the maximum speed of the W band is smaller than the experimental at $n$ > 19, the computed dependence does not fit the data well. Also, around the kink, the computed $T_{max}$ values become somewhat "noisy" loosing monotonic increase with the chain length. The effect is easy to understand: when the density of W band states increases so that two W states can be found within ca. $\pm\beta_{tag}$ around the end-group state energy, the wavepacket efficiency increases, reducing the $T_{max}$ value. However, at slightly longer chain length, there will be only one W state within $\pm\beta_{tag}$ of the end-group state and the transport efficiency will drop resulting in $T_{max}$ value increase. Such fluctuations are clearly seen in the waiting-time dependence (Fig. S31). The fluctuations will continue until about three W states are found in the vicinity of the end-group state, leading to a more stable efficiency of the wavepacket formation in the W band. The $T_{max}$ values obtained for $\partial S/\partial T$ $T$-dependence also experiencing a kink at n ~15, which indicates the chain length at which the wavepacket in the W band start to dominate over the wavepacket in the Tw band (Fig. 8A, magenta dots). The group velocity determined for $T_{max}$ of $\partial S/\partial T$, 32 Å/ps, corresponds approximately to the maximum speed supported by the W band (Table 1).

No conditions were found to fit well the whole range of chain lengths using two bands, Tw and W. The transport speed for the longer chains was limited by the maximum speed of the W band, which is smaller than the experimental speed. Only by using all three bands the data can be matched well (Fig. 8B). The computed $T_{max}$ points are noisy around the kink in the $T_{max}(R)$ dependence, indicating the chain length where the regime changes, as explained for W band in Figure 8A. The magenta line starts at the chain length (~40Å) from where the transport via Ro band dominates fully.

The parameters obtained for a good match are given in the Figure 8 caption. The "noise" in the computed $T_{max}$ values forced us to start the fit from $n$ = 21. For $n$ > 25 (>42Å), the noise in $T_{max}$ subsides and the speed gravitates towards the maximum speed of the Ro band.

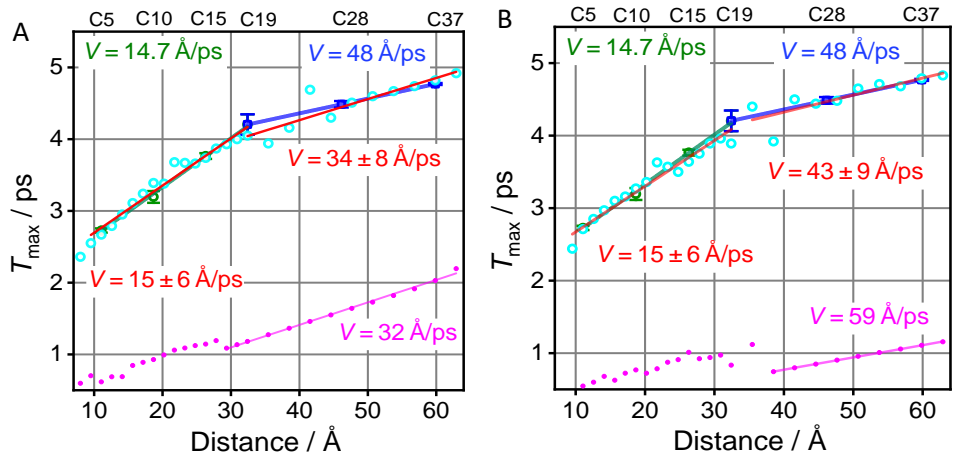

**Figure 8.** Tag-reporter distance dependence of $T_{max}$ (cyan circles) computed for the transport via (A) two chain bands, Tw and W, and (B) three chain bands, Tw, W and Ro. Linear fits of the modeling data for the $n$ of 3 – 19 and $n$ of 19-39 are shown with red lines and respective speeds are indicated as insets in red. The modeling parameters common for both panels were: $E_0$ = (-25, 25, -40) cm$^{-1}$, $\beta_{tag}$ = (12, 10, 9) cm$^{-1}$, $k_{deph}$ = 0.4 ps$^{-1}$, $k_{rlx}$ = 0.6 ps$^{-1}$, and $T_c$ = 10 ps, where the first, second, and third values in parentheses are given for the Tw, W, and Ro bands, respectively. $k_{rep}$ was at (3, 3) ps$^{-1}$ for panel A and (4.5, 3, 4) ps$^{-1}$ for panel B. The computed data (cyan circles) were shifted up by 0.6 ps (A) and by 0.65 ps (B). The overall 2-band $T$-time kinetics were obtained by summing the kinetics for individual bands with fractions of 0.15 for Tw and 0.85 for W. The overall 3-band $T$-time kinetics were obtained by summing the kinetics for individual bands with the fractions of 0.063 for Tw, 0.104 for W, and 0.833 for Ro bands. Experimental data are also shown with green and blue symbols; the lines of the matching colors show the linear fits for short and long chains, taken from Fig. 5.

Thus, the three-band model results in a reasonable explanation of the observed transport velocities. Yet there are substantial fluctuations of transport times that exceed those observed experimentally. Potentially this problem can be resolved by including anharmonic interactions with low frequency acoustic phonons as outlined below as an alternative scenario of energy transport through optical bands.

### 3.5. Effect of anharmonic interactions with acoustic phonons.

In the above consideration we ignored the effect of anharmonic interaction. In particular, anharmonic interactions with gapless acoustic phonons, inevitably present in any chain and delocalized throughout it, can be involved in optical phonon transitions between different band states. Such transitions can include backwards scattering accompanied with the overturn of the wavepacket propagation and replacing the transport mechanism from ballistic to diffusive. However, under the present condition of a sound velocity substantially exceeding the transport velocity, the scattering due to the interaction with longitudinal or torsional acoustic phonons is practically forbidden due to the Cherenkov's constraint.[45] An interaction with transverse acoustic phonons is not subject to the Cherenkov's constraint, because transverse phonons group velocity approaches zero at zero wavevector.[46] However, the bandwidth of the transverse acoustic phonons for alkyl chains, governed by interactions through covalent bonds, exceeds the bandwidth of the optical phonons. As a result, the strongest third order anharmonic interaction can lead only to the forward scattering of optical phonons, while the backward scattering requires higher order processes, which are orders of magnitude slower.[47] If the initial group velocity of optical phonon is smaller compared to its typical velocity within the band, then the fast forward scattering redistributes this phonon between the band states with the same direction of propagation that will lead to the substantial increase of the phonon transport velocity with the time. This increase is equivalent to the increase of transport velocity for longer chains as observed in the presented experiment. For example, if a wavepacket is created at the top of an optical chain band featuring higher density of states but smaller "local" group velocity, anharmonic interaction with transverse acoustic phonons can cause the wavepacket to "migrate" to the middle of the band where the group velocity is much higher, still moving in the same direction. Most optical bands of the chain can be involved in such a process, which can result in a change of the observed transport speed.

Another opportunity for changing the transport speed is associated with a phonon transition from the given optical band to another with the lower energy. This transition can be due to anharmonic interactions involving three optical phonon bands or the solvent. Following this scenario a wavepacket initiated in a twisting band can be transferred to the rocking band for longer chains. Of course the propagation direction of the transferred phonon can vary; yet around a half of them will keep propagating towards the reporter site and their further backscattering could be limited.[47]

## 4. Discussion

Azido groups offer multiple and potentially unique mechanisms of ballistic transport initiation in alkyl chains. Moreover, different mechanisms of initiation depend differently on the chain length. The transport mechanism via the Tw band involves two IVR steps; the ability to form a wavepacket in short chains ($n <$ 16) is essentially independent of the chain length. The mechanism of the transport via the W and Ro bands is inefficient at short chains as the density of these chain states is insufficient to form a wavepacket. However, with an increase of the chain length, the wavepackets at W and Ro bands became increasingly more efficient, at the same time reducing the efficiency of wavepacket formation at the Tw band. This competitive switch occurs for the chain length of ca. 16-21; the switch results in a change of the apparent energy transport speed, which becomes ca. 3-fold faster.

Modeling of the transport involving only the Tw chain band showed that the kink in the $T_{max}(R)$ dependence can be reproduced under conditions that wavepacket roundtrips contribute to the energy delivery to the reporter site. In this case, the kink occurs at chain lengths for which the roundtrip contributions become negligible. While the transport speed at small chain lengths and the kink location can be reproduced well via the modeling, the transport speed for longer chains cannot be reproduced. It is clearly shown that the apparent transport speed cannot exceed the maximal transport speed supported by the involved chain band. That is why incorporation of the W band, which starts to contribute at longer chains, cannot solve the problem either as the maximal transport speed supported by the W band is still smaller than the experimentally observed speed for longer chains. The Ro band supplies a significantly higher speed and adding a wavepacket on the Ro band enabled us to match the modeling with the experiment. We did not consider interference of different wavepackets and treated the three wavepackets as independent, as if they occur in different molecules. This approach is likely justified even if several wavepackets propagate in the same molecule because of orthogonality and different symmetry of different chain bands. Because the transport speed supported by different chain bands differ significantly, the interference effects are expected to be small as the wavepackets arrive separately to the reporter site, especially for longer chains. Note that the time of the first arrival of each wavepacket is 2-3 folds smaller than the $T_{max}$ values observed (compare the magenta line with the experimental data in Figures 7-8). That is because the observed maximum in the waiting time dependence is formed as a competition of the excess energy arriving to the end group and the end-group cooling.[36, 44]

A "noise" in the $T_{max}$ values was observed in the modeling at chain lengths for which a new band starts contributing to the energy transport, but its density of states is still small. While we cannot say with certainty that such an effect is not observed in the experiment because of a small number of chain lengths measured but all experimental waiting time dependences measured did not show multiple peaks as in Figure S31. Possible reasons for smoothing in the experimental data is in the diversity of the molecular structures of the chain, which includes deviations of the angles and distances in the chain from those corresponding to the potential energy minima for isolated chains.

**Conclusions**

It is shown that the transport speed of optical phonons propagating through alkane chains is getting faster with increasing the chain length. The observed changes of the apparent transport rate for longer chains, observed experimentally, suggested a switch of the transport regime. Modeling involving numerical solution of the Liouville equation for density matrix revealed that involvement of a single chain band (Tw) or two bands (Tw and W) cannot describe the experiment. The obtained modeling parameters, $k_{deph}$ = 0.4 ps$^{-1}$ and $k_{rlx}$ = 0.6 ps$^{-1}$, revealed that multiple roundtrips involving wavepacket reflections at the chain ends play a role only to short distances. While the wavepacket is expected to undergo some broadening and deformation due to nonlinearity of the dispersion curves for optical bands, the modeling suggests that the wavepacket can potentially exist in the chain for a long time, well exceeding 10 ps, given that the dephasing and relaxation of the chain states are slow. These observations open an opportunity of sending the wavepacket to larger distances at lower temperatures. The study provides critical testing of the mechanisms of ballistic transport in oligomeric chains. Additional mechanisms of the transport initiation appearing for longer chains are more efficient as they eliminate IVR steps in making the wavepackets in W and Ro band. The increased speed causes a decrease in losses in the chain, thus also increasing the overall transport efficiency.

**Supporting Information**
The Supporting Information includes additional figures and details of 2DIR measurements, synthesis and characterization of the test compounds, and vibrational dynamics modeling.

**Acknowledgement**
The work was supported by the National Science Foundation (CHE-2201027 to I.R. and A.B. and CHE-2145383 to N.B.).

**References**

(1) Arrhenius, S. A. Über die Dissociationswärme und den Einfluß der Temperatur auf den Dissociationsgrad der Elektrolyte. *Z. Phys. Chem.* **1889**, *4*, 96–116. DOI: 10.1515/zpch-1889-0408.
(2) Leitner, D. M. Heat Transport in Molecules and Reaction Kinetics: The Role of Quantum Energy Flow and Localization. In *Geometric Structures of Phase Space in Multidimensional Chaos*, 2005; pp 205-256.
(3) Nitzan, A. Molecules take the heat. *Science* **2007**, *317*, 759-760.
(4) Leitner, D. M. Heat transport in molecules and reaction kinetics: the role of quantum energy flow and localization. *Adv. Chem. Phys.* **2005**, *130 B*, 205-256.
(5) Cahill, D. G.; Ford, W. K.; Goodson, K. E.; Mahan, G. D.; Majumdar, A.; Maris, H. J.; Merlin, R.; S.R., P. Nanoscale thermal transport. *J. Appl. Phys.* **2003**, *93* (2), 793-818. DOI: Doi 10.1063/1.1524305.
(6) Segal, D.; Nitzan, A.; Hanggi, P. Thermal conductance through molecular wires. *J. Chem. Phys.* **2003**, *119* (13), 6840-6855. DOI: Doi 10.1063/1.1603211.
(7) Botan, V.; Backus, E. H.; Pfister, R.; Moretto, A.; Crisma, M.; Toniolo, C.; Nguyen, P. H.; Stock, G.; Hamm, P. Energy transport in peptide helices. *Proc. Natl. Acad. Sci. U.S.A.* **2007**, *104* (31), 12749-12754, Research Support, Non-U.S. Gov't. DOI: 10.1073/pnas.0701762104.
(8) Bae, M.-H.; Li, Z.; Aksamija, Z.; Martin, P. N.; Xiong, F.; Ong, Z.-Y.; Knezevic, I.; Pop, E. Ballistic to diffusive crossover of heat flow in graphene ribbons. *Nature Commun.* **2013**, *4*, 1734. DOI: doi: 10.1038/ncomms2755.
(9) Leitner, D. M.; Pandey, H. D.; Reid, K. M. Energy Transport across Interfaces in Biomolecular Systems. *J. Phys. Chem. B* **2019**, *123* (45), 9507-9524. DOI: 10.1021/acs.jpcb.9b07086.

(10) Karmakar, S.; Keshavamurthy, S. Intramolecular vibrational energy redistribution and the quantum ergodicity transition: a phase space perspective. *Physical Chemistry Chemical Physics* **2020**, *22* (20), 11139-11173, 10.1039/D0CP01413C. DOI: 10.1039/D0CP01413C.
(11) Hu, Y.; Zeng, L.; Minnich, A. J.; Dresselhaus, M. S.; Chen, G. Spectral mapping of thermal conductivity through nanoscale ballistic transport. *Nat Nanotechnol* **2015**, *10* (8), 701-706. DOI: 10.1038/nnano.2015.109  From NLM.
(12) Zaragoza, J. P. T.; Offenbacher, A. R.; Hu, S.; Gee, C. L.; Firestein, Z. M.; Minnetian, N.; Deng, Z.; Fan, F.; Iavarone, A. T.; Klinman, J. P. Temporal and spatial resolution of distal protein motions that activate hydrogen tunneling in soybean lipoxygenase. *Proceedings of the National Academy of Sciences* **2023**, *120* (10), e2211630120. DOI: doi:10.1073/pnas.2211630120.
(13) Jiang, Y.; Liu, L. C.; Sarracini, A.; Krawczyk, K. M.; Wentzell, J. S.; Lu, C.; Field, R. L.; Matar, S. F.; Gawelda, W.; Müller-Werkmeister, H. M.; Miller, R. J. D. Direct observation of nuclear reorganization driven by ultrafast spin transitions. *Nature Communications* **2020**, *11* (1), 1530. DOI: 10.1038/s41467-020-15187-y.
(14) Rubtsov, I. V.; Burin, A. L. Ballistic and diffusive vibrational energy transport in molecules. *J. Chem. Phys.* **2019**, *150* (2), 020901(020901-020918).
(15) Gotsmann, B.; Gemma, A.; Segal, D. Quantum phonon transport through channels and molecules—A Perspective. *Applied Physics Letters* **2022**, *120* (16). DOI: 10.1063/5.0088460 (acccessed 5/21/2024).
(16) Mizutani, Y.; Mizuno, M. Time-resolved spectroscopic mapping of vibrational energy flow in proteins: Understanding thermal diffusion at the nanoscale. *The Journal of chemical physics* **2022**, *157* (24). DOI: 10.1063/5.0116734 (acccessed 5/21/2024).
(17) Shen, S.; Henry, A.; Tong, J.; Zheng, R.; Chen, G. Polyethylene nanofibres with very high thermal conductivities. *Nature Nanotech.* **2010**, *5*, 251-255.
(18) Henry, A.; Chen, G. High thermal conductivity of single polyethylene chains using molecular dynamics simulations. *Phys. Rev.  Lett.* **2008**, *101*, 235502.
(19) Majumdar, S.; Sierra-Suarez, J. A.; Schiffres, S. N.; Ong, W.-L.; Higgs, C. F.; McGaughey, A. J. H.; Malen, J. A. Vibrational Mismatch of Metal Leads Controls Thermal Conductance of Self-Assembled Monolayer Junctions. *Nano Lett* **2015**, *15* (5), 2985-2991.
(20) Meier, T.; Menges, F.; Nirmalraj, P.; Hölscher, H.; Riel, H.; Gotsmann, B. Length-Dependent Thermal Transport along Molecular Chains. *Phys. Rev.  Lett.* **2014**, *113*, 060801.
(21) Sufrin, S.; Cohn, B.; Chuntonov, L. Probing the anharmonicity of vibrational polaritons with double-quantum two-dimensional infrared spectroscopy. *Nanophotonics* **2024**. DOI: doi:10.1515/nanoph-2023-0683 (acccessed 2024-05-21).
(22) Yang, Z.; Xiong, W. Molecular Vibrational Polaritons Towards Quantum Technologies. *Advanced Quantum Technologies* **2022**, *5* (8), 2100163. DOI: https://doi.org/10.1002/qute.202100163.
(23) Yue, Y.; Qasim, L. N.; Kurnosov, A. A.; Rubtsova, N. I.; Mackin, R. T.; Zhang, H.; Zhang, B.; Zhou, X.; Jayawickramarajah, J.; Burin, A. L.; Rubtsov, I. V. Band-selective ballistic energy transport in alkane oligomers: toward controlling the transport speed. *J. Phys. Chem.  B* **2015**, *119* (21), 6448-6456. DOI: 10.1021/acs.jpcb.5b03658.
(24) Schwarzer, D.; Kutne, P.; Schroeder, C.; Troe, J. Intramolecular vibrational energy redistribution in bridged azulene-anthracene compounds: Ballistic energy transport through molecular chains. *J. Chem.  Phys.* **2004**, *121* (4), 1754-1764.
(25) Schwarzer, D.; Hanisch, C.; Kutne, P.; Troe, J. Vibrational Energy Transfer in Highly Excited Bridged Azulene-Aryl Compounds: Direct Observation of Energy Flow through Aliphatic Chains and into the Solvent. *J. Phys. Chem.  A* **2002**, *106* (35), 8019-8028.

(26) Lin, Z.; Zhang, N.; Jayawickramarajah, J.; Rubtsov, I. V. Ballistic energy transport along PEG chains: distance dependence of the transport efficiency. *Phys. Chem. Chem. Phys.* **2012**, *14* (30), 10445-10454. DOI: 10.1039/c2cp40187h.

(27) Kurochkin, D. V.; Naraharisetty, S. R.; Rubtsov, I. V. A relaxation-assisted 2D IR spectroscopy method. *Proc. Natl. Acad. Sci. U.S.A.* **2007**, *104* (36), 14209-14214, Research Support, Non-U.S. Gov't. DOI: 10.1073/pnas.0700560104.

(28) Naraharisetty, S. G.; Kasyanenko, V. M.; Rubtsov, I. V. Bond connectivity measured via relaxation-assisted two-dimensional infrared spectroscopy. *J. Chem. Phys.* **2008**, *128*, 104502.

(29) Qasim, L. N.; Kurnosov, A. A.; Yue, Y.; Lin, Z.; Burin, A. L.; Rubtsov, I. V. Energy transport in PEG oligomers: Contributions of different optical bands. *J. Phys. Chem. C* **2016**, *120* (47), 26663-26677. DOI: DOI: 10.1021/acs.jpcc.6b09389.

(30) Mackin, R. T.; Leong, T. X.; Rubtsova, N. I.; Burin, A. L.; Rubtsov, I. V. Low-Temperature Vibrational Energy Transport via PEG Chains. *J. Phys. Chem. Lett.* **2020**, *11* (12), 4578-4583. DOI: 10.1021/acs.jpclett.0c01273.

(31) Lin, Z.; Rubtsov, I. V. Constant-speed vibrational signaling along polyethyleneglycol chain up to 60-Å distance. *Proc. Natl. Acad. Sci. U.S.A.* **2012**, *109* (5), 1413-1418. DOI: 10.1073/pnas.1116289109.

(32) Rubtsova, N. I.; Rubtsov, I. V. Ballistic energy transport via perfluoroalkane linkers. *Chem. Phys.* **2013**, *422*, 16-21. DOI: 10.1016/j.chemphys.2013.01.026.

(33) Rubtsova, N. I.; Qasim, L., N.; Kurnosov, A. A.; Burin, A. L.; Rubtsov, I. V. Ballistic energy transport in oligomers. *Acc. Chem. Res.* **2015**, *48*, 2547-2555. DOI: 10.1021/acs.accounts.5b00299.

(34) Qasim, L. N.; Atuk, E. B.; Maksymov, A. O.; Jayawickramarajah, J.; Burin, A. L.; Rubtsov, I. V. Ballistic Transport of Vibrational Energy through an Amide Group Bridging Alkyl Chains. *J. Phys. Chem. C* **2019**, *123*, 3381-3392. DOI: 10.1021/acs.jpcc.8b11570.

(35) Leong, T. X.; Qasim, L. N.; Mackin, R. T.; Du, Y.; Jr., R. A. P.; Rubtsov, I. V. Unidirectional coherent energy transport via conjugated oligo(p-phenylene) chains. *J. Chem. Phys.* **2021**, *154* (13), 134304. DOI: 10.1063/5.0046932.

(36) Nawagamuwage, S. U.; Qasim, L. N.; Zhou, X.; Leong, T. X.; Parshin, I. V.; Jayawickramarajah, J.; Burin, A. L.; Rubtsov, I. V. Competition of Several Energy-Transport Initiation Mechanisms Defines the Ballistic Transport Speed. *J. Phys. Chem. B* **2021**, *125* (27), 7546-7555. DOI: 10.1021/acs.jpcb.1c03986.

(37) Varner, C.; Zhou, X.; Saxman, Z. K.; Leger, J. D.; Jayawickramarajah, J.; Rubtsov, I. V. Azido alkanes as convenient reporters for mobility within lipid membranes. *Chem. Phys.* **2018**, *512*, 20-26. DOI: doi: 10.1016/j.chemphys.2018.05.020.

(38) Rubtsov, I. V. Relaxation-assisted 2DIR: Accessing distances over 10Å and measuring bond connectivity patterns. *Acc. Chem. Res.* **2009**, *42* (9), 1385-1394.

(39) Leger, J. D.; Nyby, C. M.; Varner, C.; Tang, J.; Rubtsova, N. I.; Yue, Y.; Kireev, V. V.; Burtsev, V. D.; Qasim, L. N.; Rubtsov, G. I. Fully automated dual-frequency three-pulse-echo 2DIR spectrometer accessing spectral range from 800 to 4000 wavenumbers. *Review of Scientific Instruments* **2014**, *85* (8).

(40) Nyby, C. M.; Leger, J. D.; Tang, J.; Varner, C.; Kireev, V. V.; Rubtsov, I. V. Mid-IR beam direction stabilization scheme for vibrational spectroscopy, including dual-frequency 2DIR. *Opt. Express* **2014**, *22* (6), 6801-6809. DOI: 10.1364/OE.22.006801.

(41) Kurnosov, A. A.; Rubtsov, I. V.; Burin, A. L. Fast transport and relaxation of vibrational energy in polymer chains. *J. Chem. Phys.* **2015**, *142* (1), 011101.

(42) Borrelli, R.; Gelin, M. F. Quantum dynamics of vibrational energy flow in oscillator chains driven by anharmonic interactions. *New Journal of Physics* **2020**, *22* (12), 123002. DOI: 10.1088/1367-2630/abc9ed.

(43) Fayer, M. D. *Elements of quantum mechanics*; Oxford University Press, 2001.

(44) Rubtsova, N. I.; Lin, Z.; Mackin, R. T.; Rubtsov, I. V. How Intramolecular Vibrational Energy Transport Changes with Rigidity and Polarity of the Environment? *High Energy Chem.* **2020**, *54* (6), 427–435.

(45) Burin, A. L.; Parshin, I. V.; Rubtsov, I. V. Maximum propagation speed and Cherenkov effect in optical phonon transport through periodic molecular chains. *The Journal of chemical physics* **2023**, *159* (5). DOI: 10.1063/5.0158201 (acccessed 5/11/2024).

(46) Boulatov, A.; Burin, A. L. Crucial effect of transverse vibrations on the transport through polymer chains. *J. Chem. Phys.* **2020**, *153* (13), 134102. DOI: 10.1063/5.0018591.

(47) Burin, A. L.; Rubtsov, I. V. Two stage decoherence of optical phonons in periodic oligomers. *To be submitted*.

**Supporting Information**

**Ballistic Energy Transport via Long Alkyl Chains: A New Initiation Mechanism**

Sithara U. Nawagamuwage, Elliot S. Williams, Md Muhaiminul Islam, Igor V. Parshin, Alexander L. Burin, Nathalie Busschaert, Igor V. Rubtsov*

*Department of Chemistry, Tulane University, New Orleans, Louisiana 70118, United States*

irubtsov@tulane.edu

**S1.** <u>Synthesis</u>

**S1.1. Reaction Schemes**

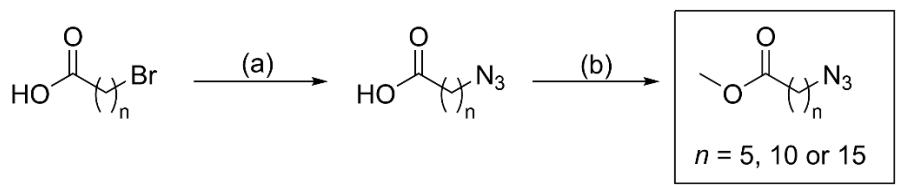

**Scheme S1.** Short azide fatty esters reaction scheme. The shortest chain was sourced as ω-azidofatty acid chain. Az5ME (n = 5, 99% yield), az10ME (n = 10, 77% yield), and az15ME (n = 15, 86% yield). (a) NaN$_3$ (4 eq.), DMF, 20 h, r.t. (b) dilute MeOH/HCl, toluene/MeOH, 50 °C, 18 h.

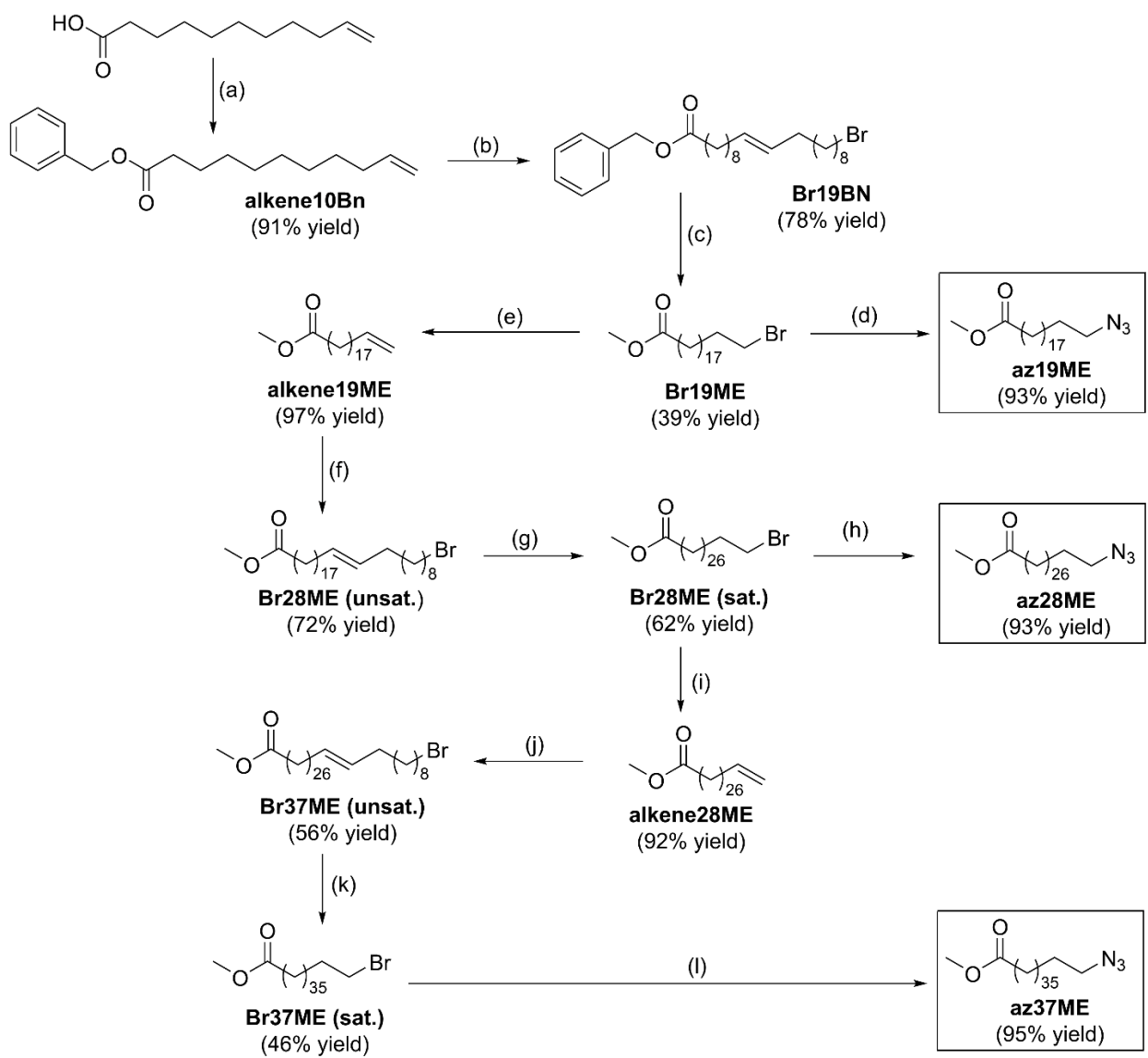

**Scheme S2.** Long azide fatty esters reaction scheme: (a) 1. NaOH, MeOH, 30 min, r.t., 2. BnBr (1.2 eq), DMF, 90 °C, 18 h. (b) 11-Bromoundecene (10 eq.), 40 °C, 20 h. (c) $H_2$, 1:2 MeOH/ethyl acetate, Pd/C cat. (10 wt. %). (d) 4 eq. $NaN_3$, DMF, r.t. 20 h (e) $Pd_2(dba)_3$, [HP(*t*-Bu)$_2$Me)]BF$_4$, dioxane, *t*-BuOK, Cy$_2$NH. (f) 11-Bromoundecene (10 eq.), DCM, 40 °C, 20 h. (g) $H_2$, 1:2 MeOH/ethyl acetate, Pd/C cat. (10 wt. %) (h) 4 eq. $NaN_3$, DMF, r.t., 20 h. (i) $Pd_2(dba)_3$, [HP(*t*-Bu)$_2$Me)]BF$_4$, dioxane, *t*-BuOK, Cy$_2$NH. (j) 11-Bromoundecene (20 eq.) DCM, 40 °C, 20 h. (k) $H_2$, 1:2 MeOH/ethyl acetate, Pd/C cat. (10 wt. %). (l) 72 eq. $NaN_3$, DMF, r.t. 72 h.

## S1.2. Synthesis Procedures

S1.3. Compound names are those generated by ChemDraw 16.0.1.4 following IUPAC nomenclature. Solvents, reagents, and inorganic salt were supplied from AA Blocks, Abachemscene, Sigma-Aldrich, TCI America, and Combi-Blocks. Reagents were used without further purification unless specified. Inert reactions were conducted either under an ultra-high purity argon or ultra-high purity nitrogen atmosphere in oven-dried glassware. Flash column chromatography was carried out using SiliaFlash P60 (40-63 µm, 230-400 mesh) or basic alumina. Thin-layer chromatography was carried out using silica gel TLC plates with fluorescent indicators, visualized under UV light (254 nm), or stained with iodine, ninhydrin, or permanganate solutions. $^1$H NMR and $^{13}$C NMR, spectra were collected on either a Bruker 500 MHz NMR, Bruker Avance Neo 400 MHz NMR, Varian Unity Inova 400 MHz NMR, or a Bruker Avance 300 MHz NMR spectrometer. $^{13}$C NMR spectra were collected proton decoupled and visualized in Mestrenova. Chemical shifts (δ) are reported in parts per million (ppm) and calibrated to the residual solvent peak in CDCl$_3$ δ = 7.26 ($^1$H NMR) and 77.2 ppm ($^{13}$C NMR) or DMSO-$d_6$ δ = 2.50 ($^1$H NMR) and 39.5 ppm ($^{13}$C NMR). Coupling constants ($J$) are reported in Hertz (Hz) and rounded to the nearest 0.5 Hz. The following abbreviations are used for spin multiplicity: s = singlet, d = doublet, t = triplet, q = quartet, p = pentet, ddt = doublet of doublet of triplets, m = multiplet, and br. = broad. Infrared (IR) spectra were recorded on a Nexus 670 Avatar FTIR spectrometer; only selected maximum peaks ($v_{max}$, cm$^{-1}$) are reported. Electron spray ionization (ESI) mass spectra were recorded on a Bruker MicroTOF.

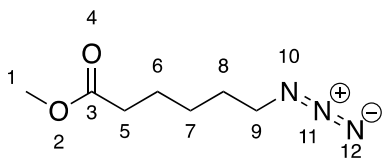

**Methyl 6-azidohexanoate,v (az5ME).** 6-Azidohexanoic acid (298 mg, 1.896 mmol) was dissolved in 2 mL toluene, 20 mL methanol, and 3 mL of dilute MeOH/HCl solution (made by adding 10 mL concentrated HCl to 40 mL of MeOH). The solution was heated to 50 °C and stirred overnight for 20 h. The reaction was sealed with a glass stopper to prevent the solvent from evaporating. After 20 h, the reaction was cooled to room temperature, and the solvent volume was reduced using the roto-evaporator leaving the crude and some toluene. The crude was dissolved in 20 mL hexanes and transferred to a 150 mL extraction funnel. 20 mL of water was also added to extraction funnel. The product was extracted into the hexanes layer and collected in a beaker. The extraction was repeated twice more with 20 mL of hexanes. The collected organic layer was dried over MgSO$_4$ and filtered to remove the drying agent. The dried hexanes solution was roto-evaporated to remove the solvent, leaving behind a clear oil. The oil was further evaporated with a gentle stream of argon gas for 30 minutes. The product is a volatile clear oil that should not be placed under high vacuum. (320 mg, 1.87 mmol, 99% yield). The spectra were in agreement with literature reported data.[1] $^1$H NMR (400 MHz, CDCl$_3$, ppm) δ 3.66 (3H, s, **1**), 3.26 (2H, t, $J$=7.0 Hz, **9**), 2.32 (2H, t, $J$=7.5 Hz, **5**), 1.70 – 1.55 (2H, m, **6, 8**), 1.47 – 1.31 (2H, m, **7**). $^{13}$C NMR (101 MHz, CDCl$_3$, ppm) δ 174.0 (**3**), 51.6 (**1**), 51.3 (**9**), 33.9 (**5**), 28.7 (**7**), 26.3 (**8**), 24.5 (**6**). IR (neat): $v$ (cm$^{-1}$) = 2946, 2865, 2091, 1735, 1346, 1352, 1251, 1001, 1173, 1154, 1098, 1009.

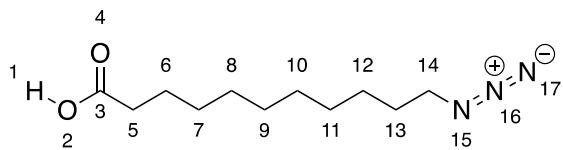

**11-Azidoundecanoic acid, (az10OH).** 11-Bromoundecanoic acid (1.361g, 5.132 mmol, 1 eq) and NaN$_3$ (1.465 g, 22.538 mmol, 4.4 eq.) were dissolved in 25 mL DMF in a 100 mL round bottom flask equipped with a magnetic stir bar. The reaction was stirred at r.t. for 20 h under ambient atmosphere. Afterwards, the solution was transferred into a 250 mL extraction funnel. 50 mL ethyl acetate was added to the round bottom flask and then transferred into the extraction funnel ensuring efficient transfer. 100 mL H$_2$O was added to the extraction funnel. The solution was then acidified with 5% HCl v/v solution. The aqueous layer was collected, and the organic layer was washed two more times with 100 mL dilute HCl solution. The organic layer was collected into a 100 mL round bottom flask, the solvent was removed via roto-evaporation, and then further dried under high vacuum. The product was obtained as a clear oil that required no further purification. (1.116 g, 4.910 mmol, 96% yield). $^1$H NMR (300 MHz, CDCl$_3$, ppm) δ 3.25 (2H, t, *J*=7.0 Hz, **16**), 2.35 (2H, t, *J*=7.5 Hz, **5**), 1.81 – 1.47 (4H, m, **6, 15**), 1.45 – 1.11 (12H, m, **7, 8, 9, 10, 11, 12, 13, 14**) The spectral features were in agreement with literature reported values.[2]

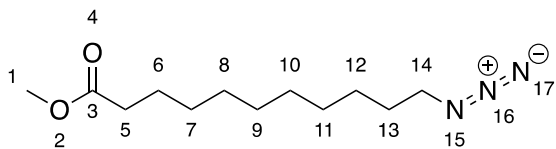

**Methyl 11-azidoundecanoate, (az10ME).** 11-azidoundecanoic acid (**az10OH**) (1.116 g, 4.901 mmol) was dissolved in 2 mL toluene, 20 mL methanol, and 3 mL of dilute MeOH/HCl solution (made by adding 10 mL concentrated HCl to 40 mL of MeOH). The solution was heated to 50 °C and stirred overnight for 18 h. The reaction was not sealed so the solvent mostly evaporated. 15 mL MeOH was added to increase the volume and then transferred into a 150 mL extraction funnel. The product was extracted 2x with 25 mL hexanes. The hexane layer was transferred to a round bottom flask, evaporated via roto-evaporation and further dried under high vacuum. The product was a pure clear oil that required no further purification (946 mg, 3.91 mmol, 80% yield). The spectra were in agreement with literature reported data.[2] $^1$H NMR (400 MHz, CDCl$_3$, ppm) δ 3.75 – 3.55 (3H, s, **1**), 3.24 (2H, t, *J*=7.0 Hz, **14**), 2.29 (2H, t, *J*=7.5 Hz, **5**), 1.68 – 1.50 (4H, m, **6, 13**), 1.27 (12H, s, **7, 8, 9, 10, 11, 12**). $^{13}$C NMR (101MHz, CDCl$_3$, ppm) δ 174.4 (**2**), 51.6 (**14**), 51.5 (**1**), 34.2 (**5**), 29.7 – 28.8 (m, **7, 8, 9, 10, 11, 12**), 26.8 (**13**), 25.0 (**6**) IR (neat): *ν* (cm$^{-1}$) = 2926, 2855, 2092, 1738, 1457, 1436, 1251, 1196, 1170, 1103.

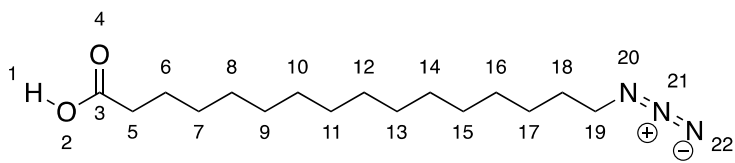

**16-Azidohexadecanoic acid, (az15OH).** 16-Bromohexadecanoic acid (465 mg, 1.386 mmol, 1 eq) and NaN$_3$ (360 mg, 5.538 mmol, 4 eq) were dissolved in 25 mL DMF in a 100 mL round bottom flask with a stir bar. The reaction was stirred at room temperature for 20 h. After 20 h the reaction was worked up by dissolving the product in ethyl acetate and washing with water to remove DMF and NaN$_3$. The reaction was incomplete according to NMR, so the crude was redissolved in 25 mL of DMF and an additional 4 eq. NaN$_3$ was added to the flask. The reaction was stirred for an additional 20 h at room temperature. Afterwards, the solution was transferred into a 250 mL extraction funnel. 50 mL ethyl acetate was added to the round bottom and then transferred into the extraction funnel. 100 mL H$_2$O was added to the extraction funnel. The solution was then acidified with 5% HCl v/v solution. The aqueous layer was drained and repeated two more times with 100 mL dilute HCl solution. The organic layer was collected into a 100 mL round bottom flask and the solvent was removed via roto-evaporation, and then further dried under high vacuum. The product was obtained as a tan solid that required no further purification. NMR spectra were in agreement with those previously reported in literature.[3] (392 mg, 1.32 mmol, 96% yield) $^1$H NMR (300 MHz, CDCl$_3$, ppm) δ 3.2 (2H, t, *J*=7.0 Hz, **5**), 2.3 (2H, t, *J*=7.5 Hz, **19**), 1.6 (4H, m, **6, 18**), 1.3 (22H, s, **7, 8, 9, 10, 11, 12, 13, 14, 15, 16, 17**). [13]C NMR (75 MHz, CDCl$_3$, ppm) δ 180.4 (**3**), 51.6 (**19**), 34.2 (**5**), 29.9 – 28.7 (m, **7, 8, 9, 10, 11, 12, 13, 14, 15, 16, 17**), 26.9 (**18**), 24.8 (**6**)

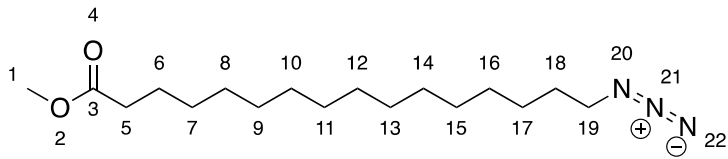

**Methyl 16-azidohexadecanoate, (az15ME).** 16-Azidohexadecanoic acid (**az15OH**) (392 mg, 1.32 mmol) was dissolved in 2 mL toluene, 20 mL methanol, and 3 mL of dilute MeOH/HCl solution (made by adding 10 mL concentrated HCl to 40 mL of MeOH). The solution was heated to 50 °C and stirred overnight for 18 h. The reaction was sealed with a glass stopper to prevent the solvent from evaporating. The solution was cooled to room temperature and transferred into a 150 mL extraction funnel. The product was extracted with 50 mL hexanes split between two additions. The hexane layers were evaporated via roto-evaporation, and further dried under high vacuum. The product obtained was a pure white solid that required no further purification (373 mg, 1.20 mmol, 90% yield) [1]H NMR (300 MHz, CDCl$_3$, ppm) δ 3.66 (s, 3H, **1**), 3.25 (t, *J* = 7.0 Hz, 2H, **19**), 2.30 (t, *J* = 7.5 Hz, 2H, **5**), 1.73 – 1.56 (m, 4H, **6, 18**), 1.44 – 1.13 (m, 22H, **7, 8, 9, 10, 11, 12, 13, 14, 15, 16, 17**). [13]C NMR (101 MHz, CDCl$_3$): 174.5 (**3**), 51.6 (**1**), 51.6 (**19**), 34.3 (**5**), 29.9 – 28.9 (m, **7, 8, 9, 10, 11, 12, 13, 14, 15, 16, 17**), 26.9 (**18**), 25.1 (**6**).). IR (neat): ν (cm$^{-1}$) = 2923, 2853, 2093, 1739, 1457, 1436, 1351, 1251, 1197, 1169, 1111, 1015.

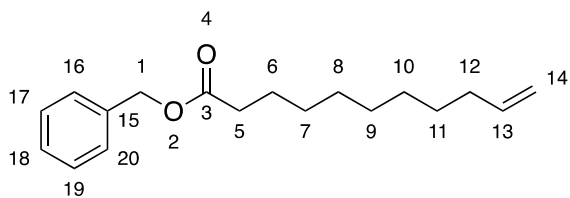

**Benzyl undec-10-enoate, (alkene10BN).** NaOH (5.13 g, 0.128 mol, 1.0 eq.) was dissolved in 70 mL MeOH. Once fully dissolved, undec-10-enoic acid (23.62 g, 0.128 mol, 1.0 eq) was slowly added to the methanol solution while stirring. A precipitate had formed after 40 mins of stirring. MeOH was removed via roto-evaporation and the flask was dried on the high vacuum for one hour. Benzyl bromide (23.229 g, 0.135 mol, 1.1 eq) was dissolved in 200 mL DMF. This solution was transferred into the sodium salt of the fatty acid and the solution was heated and stirred at 90 °C for 18 h. After 18 h, the DMF solution was transferred to a 500 mL extraction funnel, 100 mL ethyl acetate was added and washed with 100 mL DI H$_2$O (three times). The organic layer was transferred into a round bottom flask and the solvent was removed via roto-evaporation and dried on high vacuum. The crude product was then purified with a silica column using hexanes as mobile phase. The purified product was dried on high vacuum. In the case of excess benzyl bromide: the product was treated with triethylamine (0.4 eq) for 24 h with stirring. The resultant TEA-Br salt was filtered off leaving a pure clear liquid product. (32.15 g, 0.117 mols, 91% yield) [1]H NMR (300 MHz, CDCl$_3$, ppm) δ 7.46 – 7.30 (5H, m, **16,17,18,19,20**), 5.83 (1H, ddt, *J*=17.0, 10.0, 6.5 Hz, **13**), 5.14 (2H, s, **1**), 5.08 – 4.86 (2H, m, **14**), 2.38 (2H, t, *J*=7.5 Hz, **5**), 2.08 (2H, q, *J*=7.0 Hz, **12**), 1.77 – 1.55 (2H, m, **6**), 1.42 – 1.23 (10H, m, **7,8,9,10,11**). [13]C NMR (101 MHz, CDCl$_3$, ppm) δ 173.6 (**3**), 139.2 (**13**), 136.2 (**15**), 128.6 (**17, 19**), 128.3 – 128.1 (m, **16, 18, 20**), 114.2 (**14**), 66.1 (**1**), 34.4 (**5**), 33.8 (**12**), 29.5 – 28.7 (m, **7, 8, 9, 10, 11**), 25.0 (**6**). IR (neat): ν (cm$^{-1}$) = 3065, 3032, 2926, 2854, 1735, 1640, 1455, 1158, 993, 908. HRMS (ESI+) for C$_{18}$H$_{26}$O$_2$ [M+H]$^+$: *m/z* = 275.1988 (observed), 275.2011 (calculated).

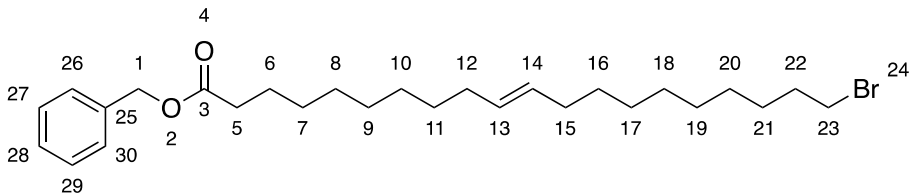

**Benzyl (*E*)-20-bromoicos-10-enoate, (Br19BN)**. Benzyl undec-10-enoate (**alkene10BN**) (25.1 g, 91.5 mmol, 1.0 eq.) was added to 11-bromoundec-1-ene (213.0 g, 913 mmol, 10.0 eq) into a 500 mL round bottom flask. This reaction was performed neat, without the use of solvents. The flask was evacuated, backfilled with argon gas, and subjected to 3 freeze-thaw cycles. While freezing, the flask was evacuated. While thawing, the flask was backfilled with argon gas. Once thawed, 2.0 g of Grubbs second generation catalyst was added and a reflux condenser with a vacuum stopcock adaptor was attached to the round bottom flask. The flask was evacuated again, backfilled with argon gas, and subjected to one more freeze thaw cycle. Once thawed, the round bottom flask was heated to 111 °C and refluxed overnight. After 20 h, the round bottom flask was removed from heat and allowed to cool. The crude oil was loaded neat onto a large silica column with a hexane mobile phase. The 11-bromoundec-1-ene starting material was eluted with hexanes. The metathesis product eluted with 10% ethyl acetate in hexanes. The pure fractions were collected in a round bottom flask and the solvent was evaporated. The product was obtained as an oil (34.0 g, 70.9 mmol 78% yield). $^1$H NMR (300 MHz, CDCl$_3$, ppm) δ 7.4 – 7.3 (5H, m, **26, 27, 28, 29, 30**), 5.5 – 5.3 (2H, m, **13, 14**), 5.1 (2H, s, **1**), 3.4 (2H, t, *J*=7.0 Hz, **23**), 2.3 (2H, t, *J*=7.5 Hz, **5**), 2.1 – 1.9 (4H, m, **12, 15**), 1.8 (2H, p, *J*=7.0 Hz, **22**), 1.6 (2H, t, *J*=7.0 Hz, **6**), 1.5 – 1.4 (2H, m, **21**), 1.4 – 1.1 (20H, m, **7, 8, 9, 10, 11, 16, 17, 18, 19, 20**). $^{13}$C NMR (101 MHz, CDCl$_3$, ppm) δ 173.8 (**3**), 136.2 (**25**), 130.4 (t, *J*=5.0 Hz, **13, 14**), 128.6 (**26, 30**), 128.3 (**27, 28, 29**), 66.2 (**1**), 34.5 (**5**), 34.2 (**23**), 33.0 (**22**), 32.7 (**12, 15**), 30.0 – 28.5 (m, **7, 8, 9, 10, 11, 16, 17, 18, 19, 20**), 28.3 (**21**), 25.1 (**6**). IR (neat): ν *(*cm$^{-1}$) = 3065, 3031, 2924, 2853, 1737, 1456, 1436, 1353, 1165, 967. HRMS (ESI+) for C$_{27}$H$_{43}$BrO$_2$ [M+H]$^+$: *m/z* = 479.2519 (calculated), 479.2527 (observed).

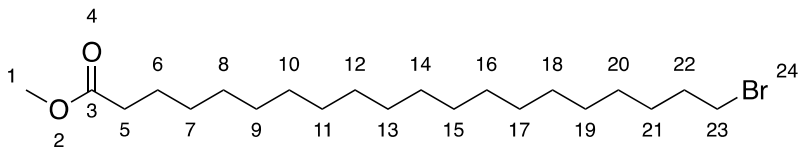

**Methyl 20-bromoicosanoate, (Br19ME)**. Benzyl 20-bromoicosanoate (**Br19BN**) (3.124 g, 6.51 mmol, 1.0 eq.) was dissolved in 30 mL 1:2 ethyl acetate:methanol and Pd/C (10 wt. %) (0.161 g, 1.51 mmol, 0.2 eq) was added to the 100 mL round bottom flask. The flask was evacuated and backfilled with argon gas (3 cycles). The flask was evacuated, and a soccer ball inflated with hydrogen gas was attached. The reaction stirred at room temperature for 20 h. After 20 h, the reaction solution was gravity filtered through a celite plug to remove Pd/C catalyst. The celite plug was washed with 50 mL hexanes to ensure the product was removed. The filtrate was evaporated leaving a crude oil, which was dissolved in methanol by heating the solution to 50 °C. Once dissolved, the solution was cooled to room temperature forming a white precipitate, and then further cooled to 8 ˚C. The precipitate was filtered off and collected with chloroform. The chloroform was evaporated off and the solid was dried under high vacuum at room temperature overnight (1.024 g, 2.53 mmol, 39 % yield). $^1$H NMR (400 MHz, CDCl$_3$, ppm) δ 3.66 (3H, s, **1**), 3.40 (2H, t, *J*=7.0 Hz, **23**), 2.30 (2H, t, *J*=7.5 Hz, **5**), 1.85 (2H, p, *J*=7.0 Hz, **22**), 1.68 – 1.57 (2H, m, **6**), 1.42 (2H, p, *J*=7.0 Hz, **21**), 1.30 – 1.21 (28H, m, **7, 8, 9, 10, 11, 12, 13, 14, 15, 16, 17, 18, 19, 20**) $^{13}$C NMR (101 MHz, CDCl$_3$, ppm) δ.5 (**3**), 51.6 (**1**), 34.3 (**5**), 34.2 (**23**), 33.0 (**22**), 29.9 – 29.1 (m, **8, 9, 10, 11, 12, 13, 14, 15, 16, 17, 18, 19, 20**), 28.9 (**7**), 28.3 (**21**), 25.1 (**6**). IR (neat): ν *(*cm$^{-1}$) = 2915, 2847, 1734, 1473, 1435, 1379, 1200, 1170, 1117, 980, 884. HRMS (ESI+) for C$_{21}$H$_{41}$BrO$_2$ [M+H]$^+$: *m/z* = 407.2350 (calculated) 407.2336 (observed).

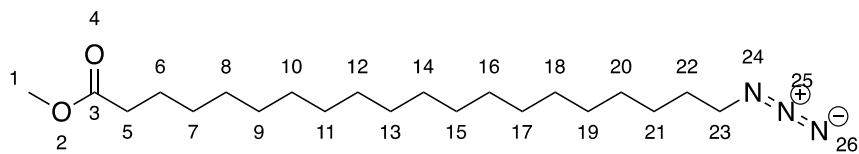

**Methyl 20-azidoicosanoate, (az19ME). Br19ME** (214 mg, 0.528 mmol, 1.0 eq), NaN$_3$ (190 mg, 2.92 mmol, 5.5 eq.) and 20 mL DMF were added to a 50 mL round bottom flask with a stir bar. The reaction was stirred at room temperature for 20 h. The reaction was then transferred to a 250 mL separatory funnel. 50 mL ethyl acetate was added and washed 3x with 100 mL DI H$_2$O to remove DMF. The organic layer was transferred to a round bottom flask and the solvent was evaporated, leaving behind a clear oil. The oil was further evaporated under a gentle stream of argon gas leaving behind **az19ME** as a white solid, which was dried overnight under high vacuum (180 mg, 0.490 mmol, 93% yield). $^1$H NMR (300 MHz, CDCl$_3$, ppm) δ 3.65 (3H, s, **1**), 3.24 (1H, t, J=7.0 Hz, **23**), 2.29 (2H, t, J=7.5 Hz, **5**), 1.69 – 1.51 (4H, m, **6, 22**), 1.24 (30H, s, **7, 8, 9, 10, 11, 12, 13, 14, 15, 16, 17, 18, 19, 20, 21**) $^{13}$C NMR (101 MHz, CDCl$_3$, ppm) δ 174.5 (**3**), 51.7 (**23**), 51.6 (**1**), 34.3 (**5**), 30.0 – 29.4 (m, **8, 9, 10, 11, 12, 13, 14, 15, 16, 17, 18, 19, 20, 21**), 29.3 (**7**), 26.9 (**22**), 25.1 (**6**). IR (neat): ν (cm$^{-1}$) = 2915, 2848, 2096, 1737, 1473, 1262, 1378, 1170, 884. HRMS (ESI+) for C$_{21}$H$_{41}$N$_3$O$_2$ [M+Na]$^+$: m/z = 390.3096 (calculated), 390.3091 (observed).

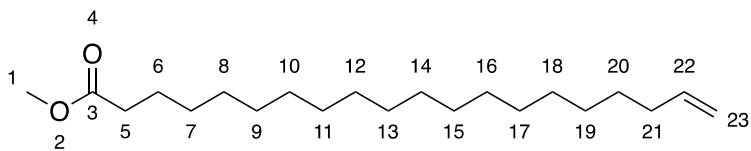

**Methyl icos-19-enoate, (alkene19ME).** Pd$_2$(dba)$_3$ (214 mg, 234 μmol, 4 mol%) [HP(t- Bu)$_2$Me)]BF$_4$ (233 mg, 803 μmol, 14 mol%), and a stir bar were added to an oven-dried 50-mL 3-neck flask. The flask was sealed with two rubber septa and a stopcock vacuum adaptor. The flask was then evacuated and backfilled with argon (3 cycles). The flask was detached from the argon line, and 10 mL dioxane (anhydrous), t-BuOK (1.5 mL of a 1 M solution in THF, 1.5 mmol, 0.25 eq.) and Cy$_2$NH (1.4 mL, 7.2 mmol, 1.2 eq.) were added sequentially via syringe. The puncture holes of the septa were sealed with vacuum grease, and the mixture was stirred at r.t. for 2 h. **Br19ME** (2.438 g, 6.01 mmol, 1.0 eq) was placed in a vial with a septum lid. The vial was purged with argon for 30 min and dissolved in 2 mL anhydrous dioxane. After 2 h of reacting the pre-catalyst, **Br19ME** was transferred into the reaction via a long needle syringe. An additional 2 mL anhydrous dioxane was used to wash the vial and ensure an efficient transfer. The reaction mixture was stirred for 24 h at room temperature. The reaction solution was then vacuumed filtered through a silica plug followed by an additional 100 mL of 5% Et$_2$O in hexanes. The filtrate was evaporated and dried on high vacuum. The product was considered pure enough to do the next reaction (1.897 g, 5.86 mmol, 97% yield). $^1$H NMR (400 MHz, CDCl$_3$, ppm) δ 5.81 (1H, ddt, J=17.0, 10.0, 7.0 Hz, **22**), 5.10 – 4.78 (2H, m, **23**), 3.66 (3H, s, **1**), 2.29 (2H, t, J=7.5 Hz, **5**), 2.03 (2H, q, J=7.0 Hz, **21**), 1.65 – 1.59 (2H, m, **6**), 1.41 – 1.17 (28H, m, **7, 8, 9, 10, 11, 12, 13, 14, 15, 16, 17, 18, 19, 20**). $^{13}$C NMR (101 MHz, CDCl$_3$, ppm) δ 174.4 (**3**), 139.4 (**22**), 114.2 (**23**), 51.6 (**1**), 34.3 (**5**), 34.0 (**21**), 30.2 – 28.4 (m, **7, 8, 9, 10, 11, 12, 13, 14, 15, 16, 17, 18, 19, 20**), 25.1 (**6**). IR (neat): ν (cm$^{-1}$) = 3065, 2915, 2848, 1739, 1641, 1466, 1435, 1365, 1166, 992, 909. HRMS (ESI+) for C$_{21}$H$_{40}$O$_2$ [M+H]$^+$: m/z = 325.3107 (calculated), 325.3079 (observed).

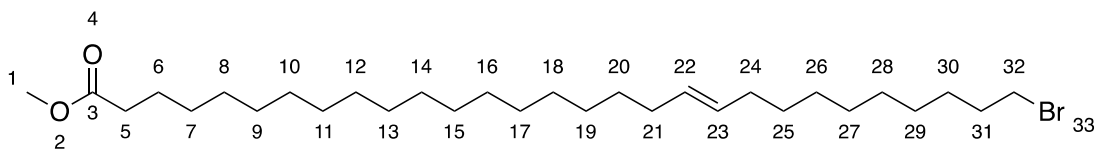

**Methyl (E)-29-bromononacos-19-enoate, (Br28ME (unsat.)).** Methyl icos-19-enoate (**alkene19ME**) (1.897 g, 5.855 mmol, 1.0 eq), 11-bromoundec-1-ene (14.26 g, 61.15 mmol, 10.44 eq), and Grubbs Second

Gen. Cat. (244 mg, 0.287 mmol, 5 mol%) were added into a 100 mL round bottom flask with a stir bar and dissolved in 40 mL DCM. A reflux condenser was attached to the flask and a vacuum adaptor stopcock was attached to the condenser. The reaction vessel was placed under high vacuum and evacuated, backfilled with argon gas, and subjected to 3 freeze-thaw cycles. While freezing, the flask was evacuated. While thawing, the flask was backfilled with argon gas. Once thawed, the round bottom flask was heated to 40 °C and refluxed overnight. After 20 h, the round bottom flask was removed from heat and allowed to cool. The crude oil was purified via column chromatography with a silica stationary phase and a gradient hexanes:ethyl acetate mobile phase. The 11-bromoundec-1-ene starting material eluted with 100% hexanes. The metathesis product eluted with 10% ethyl acetate in hexanes (2.220 g, 4.202 mmol, 72% yield). $^1$H NMR (400 MHz, CDCl$_3$, ppm) δ 5.5 – 5.3 (2H, m, **22, 23**), 3.7 (3H, s, **1**), 3.4 (2H, t, $J$=7.0 Hz, **32**), 2.3 (2H, t, $J$=7.5 Hz, **5**), 2.1 – 1.9 (4H, m, **21, 24**), 1.9 (2H, p, $J$=7.0 Hz, **31**), 1.7 – 1.6 (2H, m, **6**), 1.5 – 1.4 (2H, m, **30**), 1.4 – 1.2 (38H, m, **7, 8, 9, 10, 11, 12, 13, 14, 15, 16, 17, 18, 19, 20, 25, 26, 27, 28, 29**) $^{13}$C NMR (101 MHz, CDCl$_3$, ppm) δ 174.5 (**3**), 130.9 – 130.2 (m, **22, 23**), 51.6 (**1**), 34.3 (**32**), 34.2 (**5**), 33.0 (**31**), 32.8 (**21, 24**), 30.0 – 28.8 (m, **7, 8, 9, 10, 11, 12, 13, 14, 15, 16, 17, 18, 20, 25, 26, 27, 28, 29**), 28.3 (**30**), 25.1 (**6**). IR (neat): $\nu$ (cm$^{-1}$) = 3143, 2917, 2849, 1739, 1646, 1463, 1435, 1377, 1168, 963. HRMS (ESI+) for C$_{30}$H$_{57}$BrO$_2$ [M+Na]$^+$: $m/z$ = 553.3424 (calculated), 553.3385 (observed).

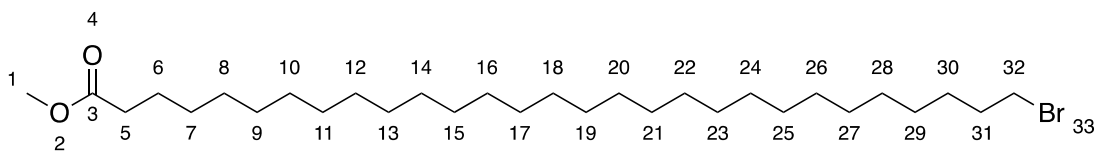

**Methyl 29-bromononacosanoate**, **(Br28ME (sat.))**. Methyl (*E*)-29-bromononacos-19-enoate (**Br28ME (unsat.)**) (645 mg, 1.22 mmol) was dissolved in 30 mL 1:1 ethyl acetate:methanol in a 100 mL round bottom flask with a magnetic stirrer. A small scoop of Pd/C (10 wt. % loading) was added to the reaction. The reaction was sealed with a vacuum adaptor equipped with a stopcock. The reaction flask was evacuated and backfilled with argon gas three times. The flask was evacuated once more and attached to a soccer ball pressurized with hydrogen gas. The reaction was stirred overnight at r.t. and monitored by TLC to ensure completion. The TLC was developed with 2% ethyl acetate in hexanes and visualized with potassium permanganate. Once complete, the solution was gravity filtered through a celite plug followed by 100 mL hexanes to collect the product. The filtrate was transferred to a round bottom flask and evaporated leaving a waxy crude. The crude was dissolved in ethyl acetate and precipitated using methanol. Once the solid precipitated, the solution was heated to 50 °C using a water bath to redissolve the solid. Once dissolved, a precipitate slowly crashed out of solution at r.t., which was removed from solution via vacuum filtration using a fine glass frit Buchner funnel. The solid was collected off the glass frit by dissolving the solid in chloroform. The chloroform solution was transferred into a round bottom flask and the solvent was removed via roto evaporation. A clear oil was obtained and further evaporated under a gentle stream of argon gas leaving behind **Br28ME (sat.)** as a pure white powder (400 mg, 0.752 mmol, 62% yield). $^1$H NMR (400 MHz, CDCl$_3$, ppm) δ 3.66 (3H, s, **1**), 3.40 (2H, t, $J$=7.0 Hz, **32**), 2.29 (2H, t, $J$=7.5 Hz, **5**), 1.85 (2H, p, $J$=7.0 Hz, **31**), 1.61 (2H, p, $J$=7.5 Hz, **6**), 1.49 – 1.36 (2H, m, **30**), 1.26 (46H, s, **7, 8, 9, 10, 11, 12, 13, 14, 15, 16, 17, 18, 19, 20, 21, 22, 23, 24, 25, 26, 27, 28, 29**). $^{13}$C NMR (101 MHz, CDCl$_3$, ppm) δ 174.4 (**3**), 51.6 , (**1**), 34.3 (**5**), 34.1 (**32**), 33.0 (**31**), 29.9 – 29.2 (m, **8, 9, 10, 11, 12, 13, 14, 15, 16, 17, 18, 19, 20, 21, 22, 23, 24, 25, 26, 27, 28, 29**), 28.9 (**7**), 28.3 (**30**), 25.1 (**6**). IR (neat): $\nu$ (cm$^{-1}$) = 2916, 2847, 1737, 1462, 1170, 975, 884, 719. HRMS (ESI+) for C$_{30}$H$_{59}$BrO$_2$ [M+H]$^+$: $m/z$ = 531.3777 (calculated) 531.3766 (observed).

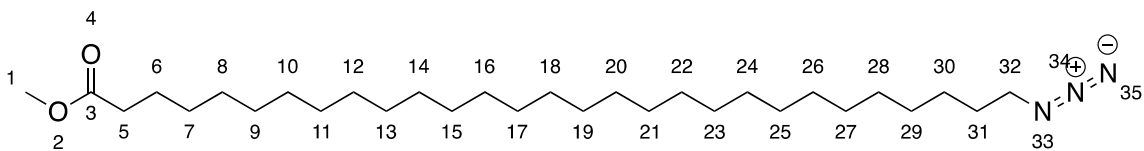

**Methyl 29-azidononacosanoate, (az28ME).** Methyl 29-bromononacosanoate (**Br28ME (sat.)**) (103 mg, 0.194 mmol, 1 eq), NaN$_3$ (186mg, 2.862 mmol, 14.7 eq), and 30 mL DMF were added to a 100 mL round bottom flask with a magnetic stir bar. The reaction was stirred at room temperature for 20 h. The product was then extracted with 50 mL ethyl acetate and washed with three additions of 100 mL DI H$_2$O to remove excess NaN$_3$. The organic layer was transferred to a round bottom flask and evaporated via roto-evaporation. The remaining solvent was evaporated by blowing a gentle stream of argon gas leaving a pure white solid of **az28ME**. The solid was dried overnight under high vacuum and required no further purification (89 mg, 180mmol, 93% yield). $^1$H NMR (400 MHz, CDCl$_3$, ppm) δ 3.66 (3H, s, **1**), 3.25 (2H, t, $J$=7.0 Hz, **32**), 2.30 (2H, t, $J$=7.5 Hz, **5**), 1.66 – 1.57 (4H, m, **6, 31**), 1.31 (48H, s, **7, 8, 9, 10, 11, 12, 13, 14, 15, 16, 17, 18, 19, 20, 21, 22, 23, 24, 25, 26, 27, 28, 29, 30**). $^{13}$C NMR (101 MHz, CDCl$_3$, ppm) δ 174.5 (**3**), 51.6 (**32**), 51.6 (**1**), 34.3 (**5**), 30.0 – 29.2 (m, **8, 9, 10, 11, 12, 13, 14, 15, 16, 17, 18, 19, 20, 21, 22, 23, 24, 25, 26, 27, 28, 29**), 29.0 (**7**), 26.9 (**31**), 25.1 (**6**). IR (neat): $\nu$ (cm$^{-1}$) = 3047, 2915, 2848, 1740, 1641, 1472, 1462, 1436, 1377, 1169, 1113, 991. HRMS (ESI+) for C$_{30}$H$_{59}$N$_3$O$_2$ [M+H]$^+$: $m/z$ = 494.4686 (calculated), 494.4695 (observed).

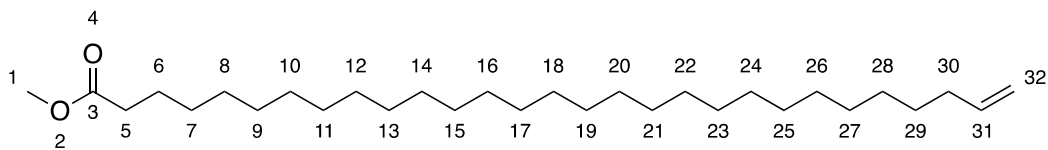

**Methyl nonacos-28-enoate, (alkene28ME).** Pd$_2$(dba)$_3$ (21 mg, 22.9 µmol, 0.16 eq.), [HP(t-Bu)$_2$Me)]BF$_4$ (21 mg, 72.3 µmol, 0.5 eq.), and a stir bar were added to an oven-dried 50 mL 3-neck flask. The flask was sealed with two rubber septa and a stopcock vacuum adaptor. The flask was then evacuated and backfilled with argon (3 cycles). The flask was detached from the argon line, and 1 mL dioxane (anhydrous), t-BuOK (150 µL of a 1 M solution in THF; 150 µmol, 1.0 eq.) and Cy$_2$NH (130 µL, 717 µmol, 4.9 eq.) were added in turn via syringe. The puncture holes of the septa were sealed with vacuum grease, and the mixture was stirred at r.t. for 2 h. **Br28ME (sat.)** (77 mg, 145 µmol, 1.0 eq.) was placed in a vial with a septum lid. The vial was purged with argon for 30 min. Once the pre-catalyst had reacted for 2 h, **Br28ME (sat.)** was dissolved in 2 mL of anhydrous dioxane, heated until fully dissolved, and transferred into the reaction via a long needle syringe. An additional 2 mL anhydrous dioxane was used to wash the vial and ensure an efficient transfer while maintaining air free conditions. The total reaction volume was 5 mL dioxane. The reaction mixture was stirred at r.t. for 28 h. The reaction solution was vacuumed filtered through a silica plug followed by an additional 100 mL 5% ether in hexanes. This filtration step removed the catalysts from the reaction mixture. The filtrate was transferred to a round bottom flask and evaporated by roto-evaporation. The crude was redissolved in hexanes and purified via silica column chromatography starting with hexanes. The product eluted in 1% ethyl acetate in hexanes. The fractions containing product were collected and combined. The solvent was evaporated using the roto-evaporator and the product was dissolved in hot ethyl acetate and precipitated overnight. The precipitate was filtered leaving a white solid. The solid was collected off the filter by dissolving it in chloroform. The chloroform was evaporated, and the product was dried under high vacuum leaving a white solid. (60 mg, 133 µmol, 92% yield). $^1$H NMR (400 MHz, CDCl$_3$, ppm) δ 5.81 (1H, ddt, $J$=17.0, 10.0, 7.0 Hz, **31**), 5.16 – 4.77 (2H, m, **32**), 3.66 (3H, s, **1**), 2.30 (2H, t, $J$=7.5 Hz, **5**), 2.04 (2H, q, $J$=7.0 Hz, **30**), 1.66 – 1.58 (2H, m, **6**), 1.41 – 1.34 (2H, m, **29**), 1.25 (44H, s, **7, 8, 9, 10, 11, 12, 13, 14, 15, 16, 17, 18, 19, 20, 21, 22, 23, 24, 25, 26, 27, 28**). $^{13}$C NMR (101 MHz, CDCl$_3$, ppm) δ 174.5 (**3**), 139.4 (**31**), 114.2 (**32**), 51.6 (**1**), 34.3 (**5**), 34.0 (**30**), 30.1 – 29.0 (m, **7, 8, 9, 10,**

**11, 12, 13, 14, 15, 16, 17, 18, 19, 20, 21, 22, 23, 24, 25, 26, 27, 28, 29**), 25.1 (**6**). IR (neat): $\nu$ (cm$^{-1}$) = 3047, 2915, 2848, 1740, 1641, 1472, 1462, 1436, 1377, 1169, 1113, 991. HRMS (ESI+) for $C_{30}H_{58}O_2$ [M+H]$^+$: $m/z$ = 451.4515 (calculated), 451.4500 (observed).

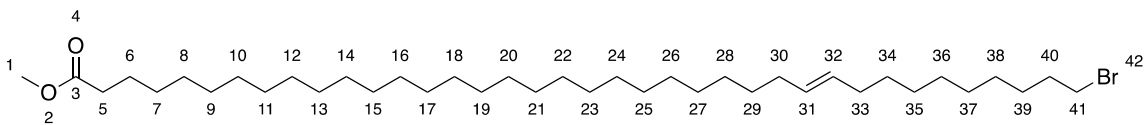

**Methyl (*E*)-38-bromooctatriacont-28-enoate, (Br37ME (unsat.))**. Methyl nonacos-28-enoate (**alkene28ME**) (54 mg, 0.120 mmol, 1 eq), 11-bromoundec-1-ene (560 mg, 2.51 mmol, 20 eq), Grubbs Catalyst 2nd Generation (8 mg, 9.42 μmol, 0.08 eq, 8 mol %), and 10 mL DCM were added to a 25 mL 3-neck round bottom flask containing a magnetic stir bar. A reflux condenser with a vacuum adaptor was connected to the round bottom flask. The remaining necks were sealed with rubber septa and wrapped with parafilm. The round bottom flask was evacuated using a high vacuum and backfilled with argon gas. The round bottom flask was then subjected to three freeze-thaw cycles to remove dissolved oxygen. While freezing, the flask was evacuated. While thawing, the flask was backfilled with argon gas. Once completely thawed, the flask was heated up to 40 °C and refluxed overnight for 28 h. The flask was cooled to r.t., the solvent was evaporated, and the contents of the round bottom flask were dried under high vacuum for 3 h to remove residual DCM, leaving behind an oily mixture. The mixture was dissolved in a minimal amount of hexanes, a dark precipitate crashed out, and the suspended solution was loaded onto a SiO$_2$ column prepared with a hexane mobile phase. The 11-bromoundec-1-ene starting material and symmetric side product were eluted with 200 mL hexanes into a single fraction. The mobile phase polarity was increased to 2% ethyl acetate in hexanes and 10 mL fractions were collected. A faint yellow band containing the product could be visualized moving down the column. Afterwards, the mobile phase was switched to DCM to collect any remaining material. The fractions containing product were collected and reduced in volume via roto-evaporation leaving behind a clear oil. The remaining solvent was removed using a stream of argon gas leaving behind a white waxy solid. The solid was then dried *in vacuo* (44 mg, 0.062 mmol, 56% yield). $^1$H NMR (400 MHz, CDCl$_3$, ppm) δ 5.47 – 5.26 (2H, m, **31, 32**), 3.66 (3H, s, **1**), 3.40 (2H, t, *J*=7.0 Hz, **41**), 2.30 (2H, t, *J*=7.5 Hz, **5**), 2.08 – 1.92 (4H, m, **30, 33**), 1.91 – 1.78 (2H, m, **40**), 1.66 – 1.59 (2H, m, **6**), 1.48 – 1.14 (58H, m, **7, 8, 9, 10, 11, 12, 13, 14, 15, 16, 17, 18, 19, 20, 21, 22, 23, 24, 25, 26, 27, 28, 29, 34, 35, 36, 37, 38, 39**). $^{13}$C NMR (101 MHz, CDCl$_3$, ppm) δ 174.5 (**3**), 130.6 (**31**), 130.4 (**32**), 51.6 (**1**), 34.3 (**5**), 34.2 (**41**), 33.0 (**40**), 32.9 – 32.4 (m, **30, 33**), 30.3 – 29.1 (m, **7, 8, 9, 10, 11, 12, 13, 14, 15, 16, 17, 18, 19, 20, 21, 22, 23, 24, 25, 26, 27, 28, 29, 34, 35, 36, 37, 38**), 28.3 (**39**), 25.1 (**6**). IR (neat): $\nu$ (cm$^{-1}$) = 3043, 2915, 2847, 1739, 1472, 1462, 1365, 1167, 964, 882, 804. HRMS (ESI+) for $C_{39}H_{75}BrO_2$ [M+Na]$^+$: $m/z$ = 679.4835 (calculated), 679.4682 (observed).

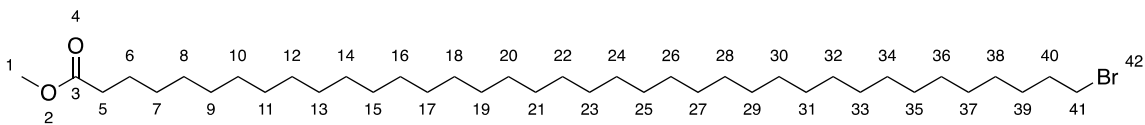

**Methyl 38-bromooctatriacontanoate, (Br37ME (sat.))**. Methyl (*E*)-38-bromooctatriacont-28-enoate (**Br37ME (unsat.)**) (37 mg, 0.056 mmol.) was added to a 100 mL round bottom flask containing a magnetic stirrer. The solid compound was dissolved in 30 mL of 1:2 MeOH:ethyl acetate. A small scoop of Pd/C (10 wt. % loading) was added to the solution and the round bottom flask was sealed with a stopcock vacuum adaptor. The flask was evacuated and backfilled with argon gas for three cycles. On the final cycle, the flask was evacuated, and a soccer ball pressurized with hydrogen gas was attached to the flask. The reaction stirred at room temperature under the hydrogen atmosphere for 18 h. The soccer ball was removed, and the solution was gravity filtered through a celite plug using hexanes. The filtrate was collected, and the solvent removed via roto-evaporation. The crude product was recrystallized by dissolving in 1 mL chloroform and precipitated using a combination of ethyl acetate and methanol. The precipitate was then redissolved by heating the flask to 50 °C in a water bath. Once redissolved, the flask was removed from the water bath and the precipitate slowly crashed out of solution at room temperature. The precipitate was vacuum filtered off using a fine glass frit Buchner funnel, washed with methanol, and collected off the frit by dissolving the solid in chloroform. The chloroform solution was reduced in volume leaving behind a clear oil. The remaining solvent was removed under a gentle stream of argon gas leaving behind a white waxy solid. The solid was then dried *in vacuo* at r.t. (17 mg, 0.025 mmol, 46% yield). [1]H NMR (400 MHz, CDCl$_3$, ppm) δ 3.66 (3H, s, **1**), 3.40 (2H, t, *J*=7.0 Hz, **41**), 2.30 (2H, t, *J*=7.5 Hz, **5**), 1.85 (2H, p, *J*=7.0 Hz, **40**), 1.66 – 1.55 (2H, m, **6**), 1.48 – 1.36 (2H, m, **39**), 1.25 (64H, s, **7, 8, 9, 10, 11, 13, 14, 15, 16, 17, 18, 19, 20, 21, 22, 23, 24, 25, 26, 27, 28, 29, 30, 31, 32, 33, 34, 35, 36, 37, 38**). [13]C NMR (101 MHz, CDCl$_3$, ppm) δ 174.5 (**3**), 51.6 (**1**), 34.3 (**5**), 34.2 (**41**), 33.0 (**40**), 30.1 – 28.8 (m, **7, 8, 9, 10, 11, 12, 13, 14, 15, 16, 17, 18, 19, 20, 21, 22, 23, 24, 25, 26, 27, 28, 29, 30, 31, 32, 33, 34, 35, 36, 37, 38**), 28.3 (**39**), 25.1 (**6**). IR (neat): ν *(*cm$^{-1}$) = 2916, 2848, 1738, 1473, 1461, 1261, 1169, 1097, 1018.

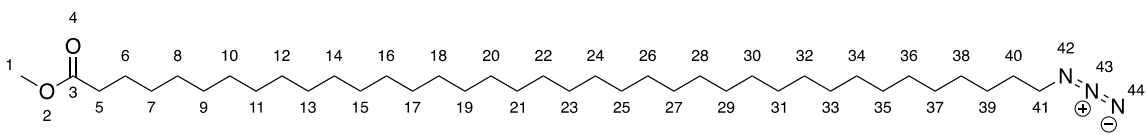

**Methyl 38-azidooctatriacontanoate, (az37ME).** Methyl 38-bromooctatriacontanoate (**Br37ME (sat.)**) (10 mg, 16 μmol, 1 eq.) and NaN$_3$ (74 mg, 1.14 mmol, 71 eq) were added to a 50 mL round bottom flask with a stir bar. The reagents were suspended in 5 mL toluene and 5 mL DMF, the flask stoppered with a rubber septum, and stirred at room temperature for 72 h. The reaction was monitored for completion using NMR. After 72 h, the solution was transferred to a 125 mL extraction funnel. 50 mL of DI H$_2$O was then added to the extraction funnel. 50 mL of ethyl acetate was used to wash the round bottom flask ensuring an efficient transfer and added to the extraction funnel. The extraction funnel was shaken, and the aqueous layer was collected. The organic layer was washed 2 more times with 50 mL DI water to remove any remaining DMF. The organic layer was collected in a pre-weighed 100 mL round bottom flask and the solvent was removed via roto-evaporation, followed by a gentle stream of argon gas, and left to dry overnight under high vacuum. The product was a white, waxy solid that required no further purification (9 mg, 15 μmol, 95% yield). [1]H NMR (400 MHz, CDCl$_3$, ppm) δ 3.66 (3H, s, **1**), 3.25 (2H, t, *J*=7.0 Hz, **41**), 2.30 (2H, t, *J*=7.5 Hz, **5**), 1.66 – 1.55 (4H, m, **6, 40**), 1.25 (66H, s, **7, 8, 9, 10, 11, 12, 13, 14, 15, 16, 17, 18, 19, 20, 21, 22, 23, 25, 26, 27, 28, 29, 30, 31, 32, 33, 34, 35, 36, 37, 38, 39**). [13]C NMR (101 MHz, CDCl$_3$, ppm) δ 174.5 (**3**), 51.7 (**41**), 51.6 (**1**), 34.3 (**5**), 30.4 – 29.1 (m, **8, 9, 10, 11, 12, 13, 14, 15, 16, 17, 18, 19, 20, 21, 22, 23, 24, 25, 26, 27, 28, 29, 30, 31, 32, 33, 34, 35, 36, 37, 38, 39**), 29.0 (**7**), 26.9 (**40**), 25.1 (**6**). IR (neat): ν *(*cm$^{-1}$) = 2916, 2848, 2096, 1472, 1462, 1365, 1143, 1094, 1018, 883, 801.

## S1.4. ¹H and ¹³C NMR spectra

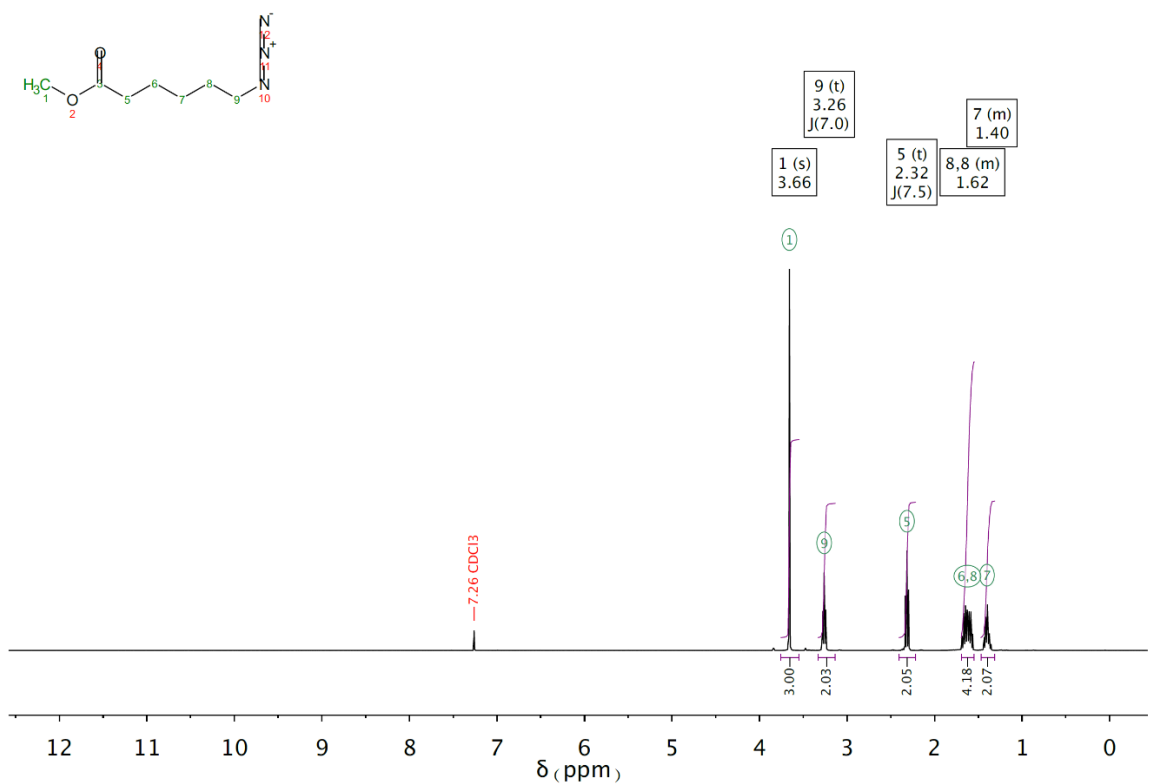

**Figure S1**. ¹H NMR (400 MHz) spectrum of **az5ME** in CDCl₃ at 298 K.

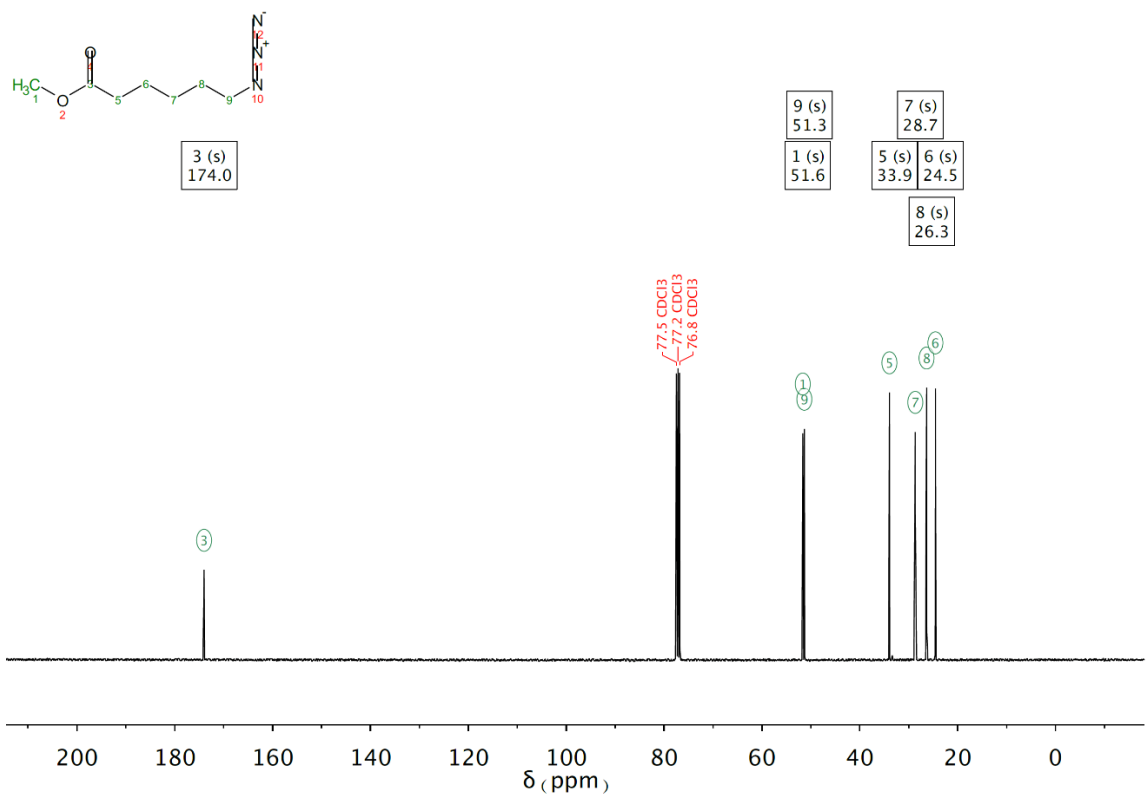

**Figure S2**. ¹³C NMR (101 MHz) spectrum of **az5ME** in CDCl₃ at 298 K.

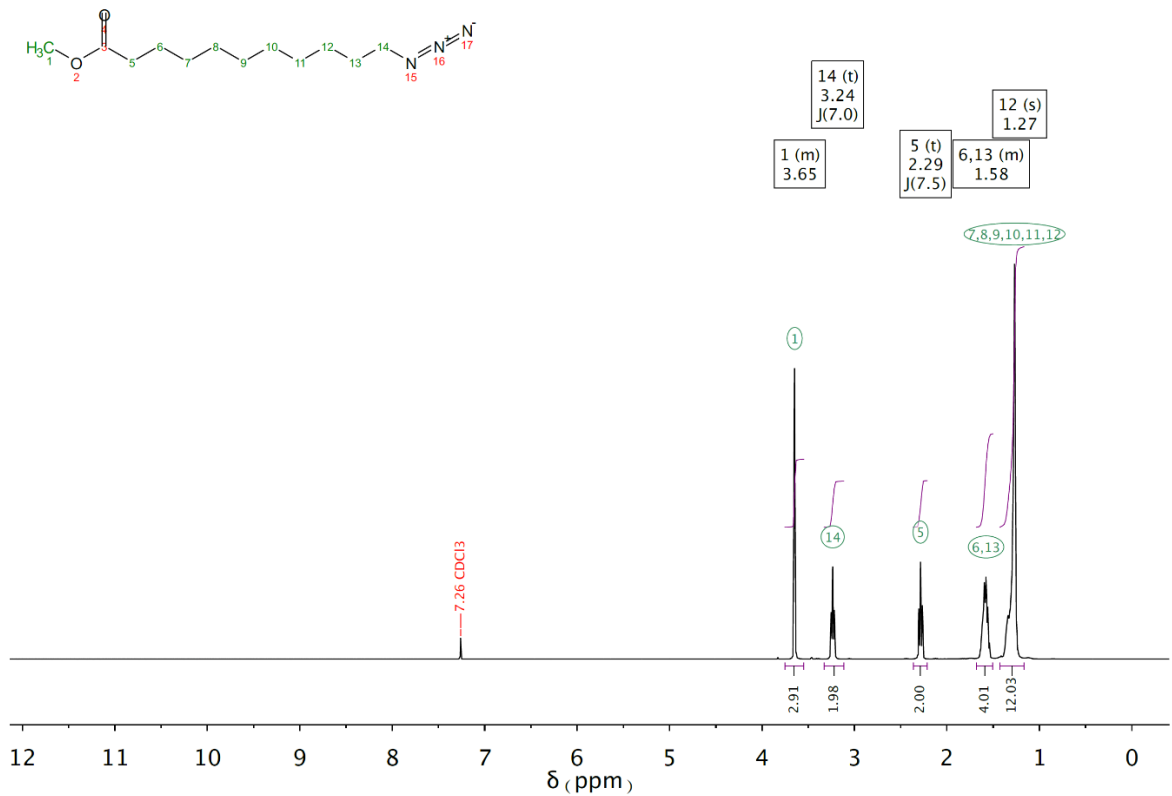

**Figure S3**. ¹H NMR (400 MHz) spectrum of **az10ME** in CDCl₃ at 298 K.

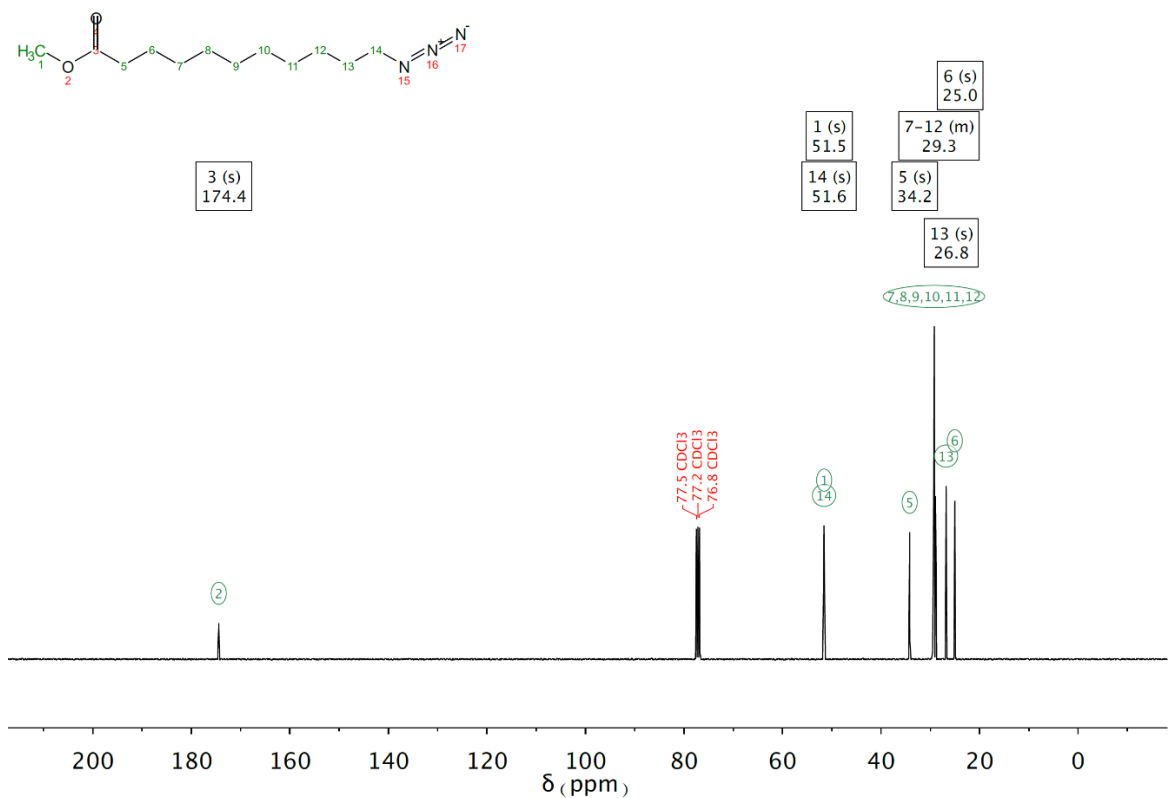

**Figure S4**. ¹³C NMR (101 MHz) spectrum of **az10OH** in CDCl₃ at 298 K.

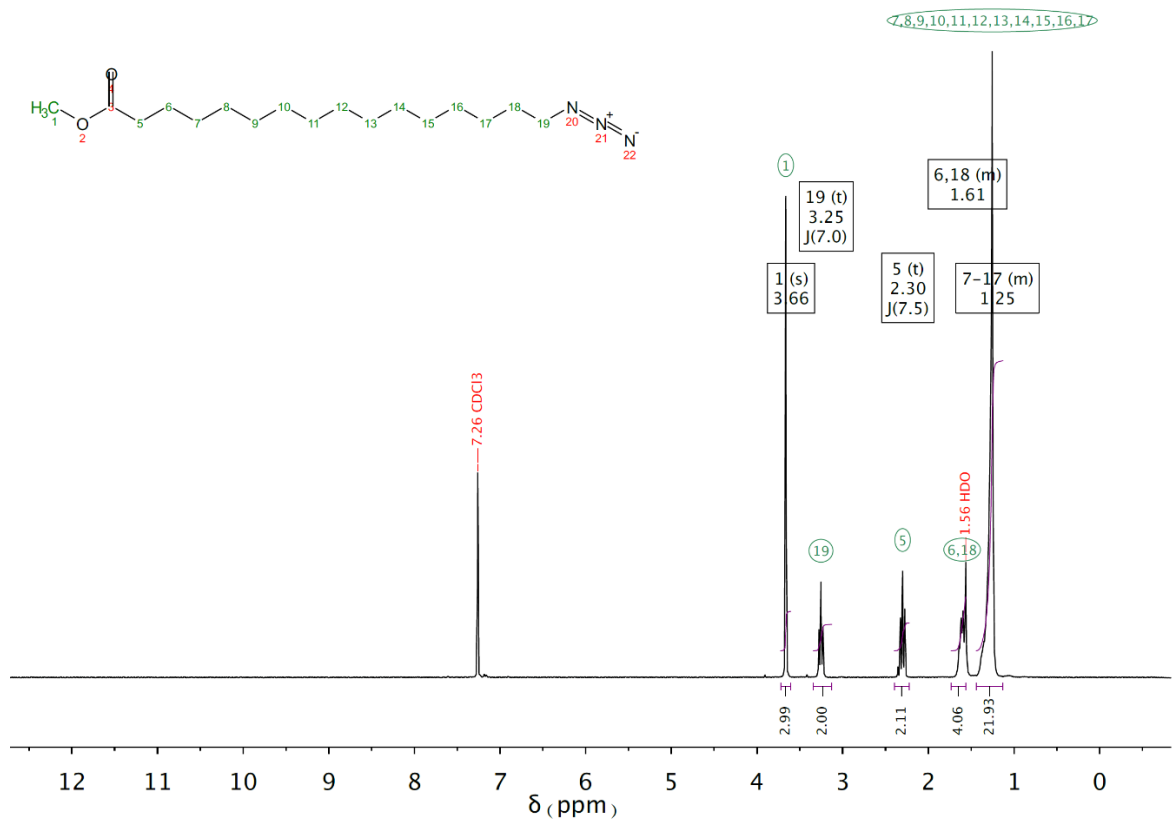

**Figure S5.** ¹H NMR (400 MHz) spectrum of **az15ME** in CDCl$_3$ at 298 K.

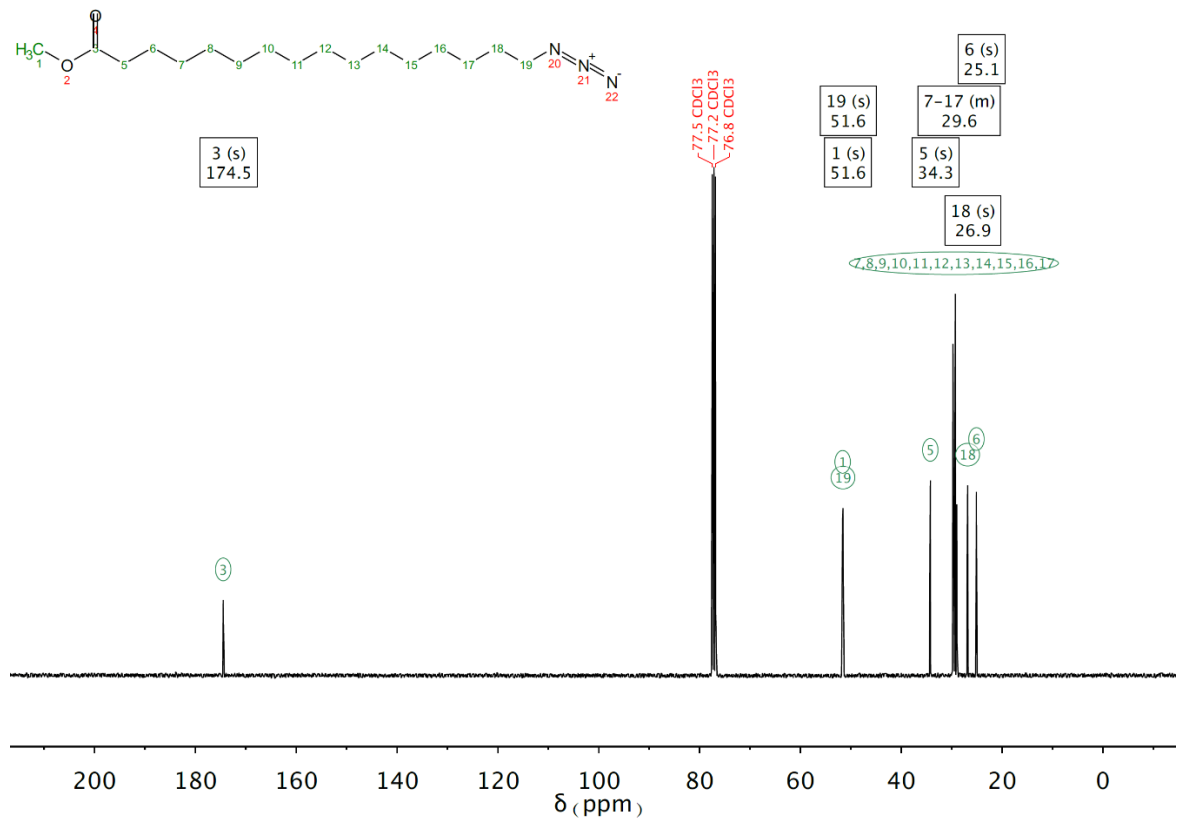

**Figure S6.** ¹³C NMR (101 MHz) spectrum of **az15ME** in CDCl$_3$ at 298 K.

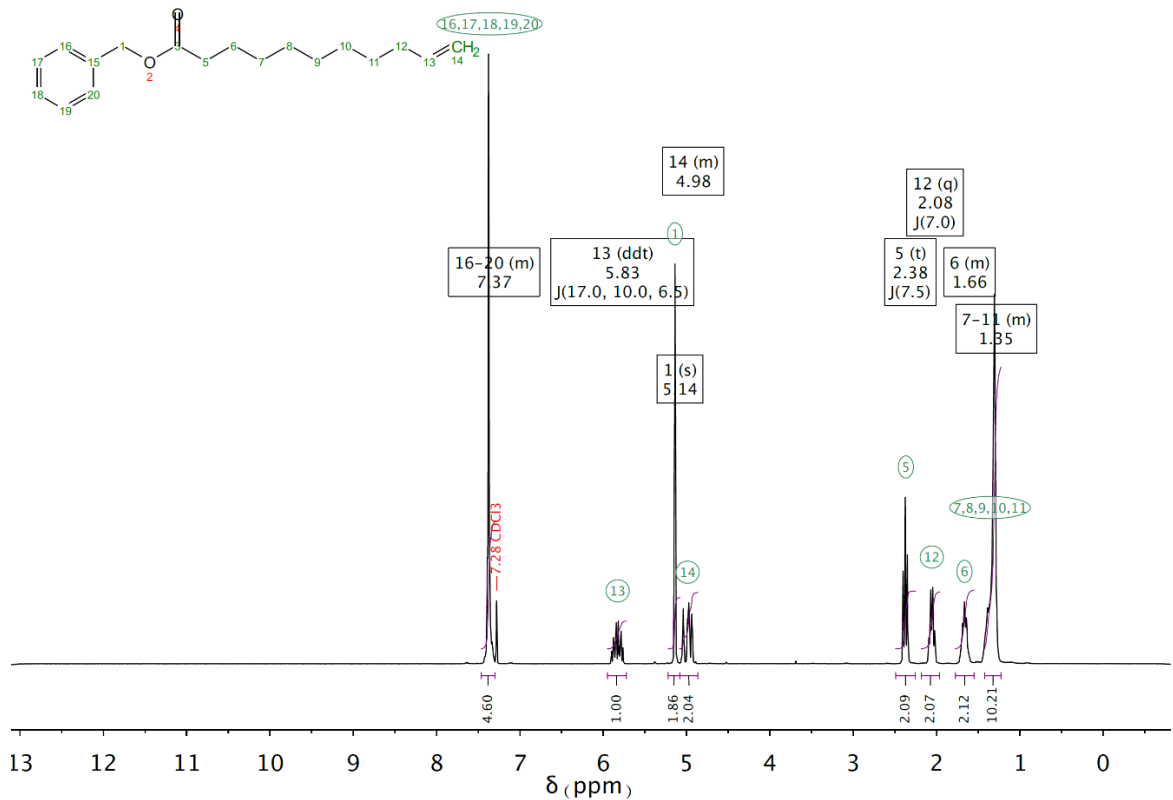

**Figure S7**. ¹H NMR (300 MHz) spectrum of **alkene10BN** in CDCl$_3$ at 298 K.

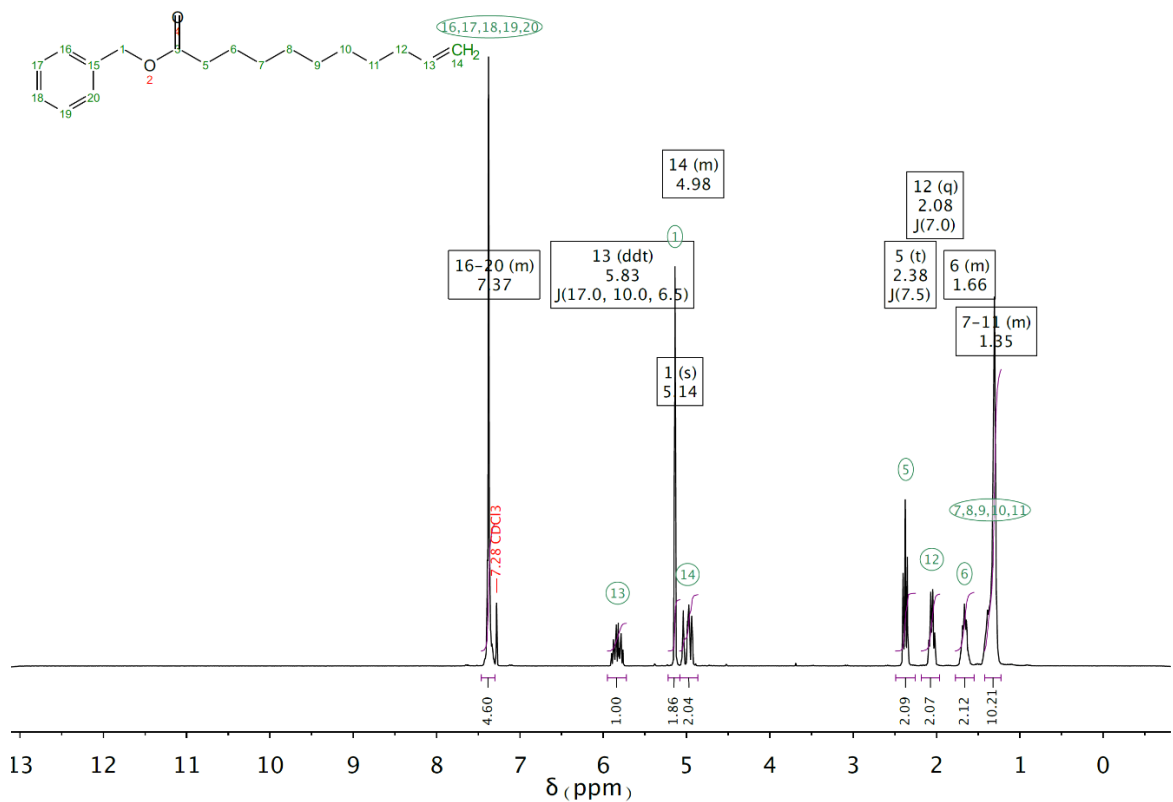

**Figure S8**. ¹³C NMR (101 MHz) spectrum of **alkene10BN** in CDCl$_3$ at 298 K.

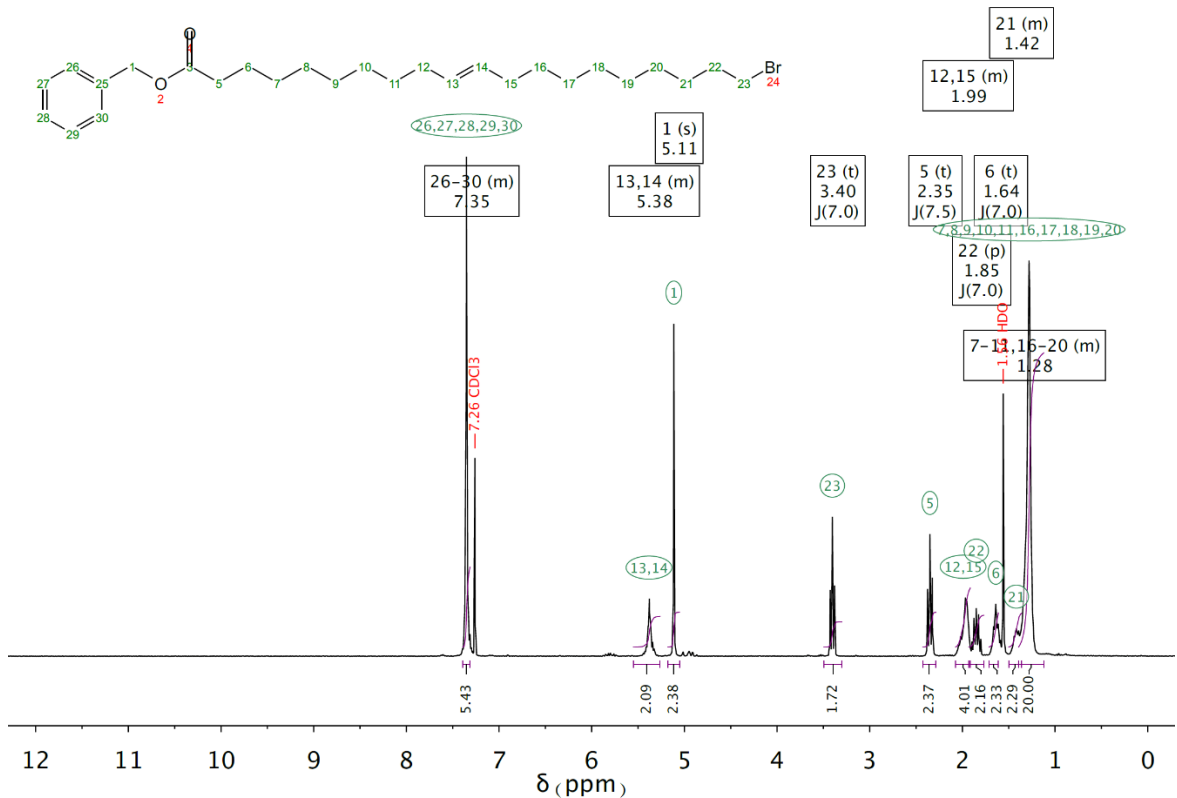

**Figure S9**. ¹H NMR (300 MHz) spectrum of **Br19BN** in CDCl₃ at 298 K.

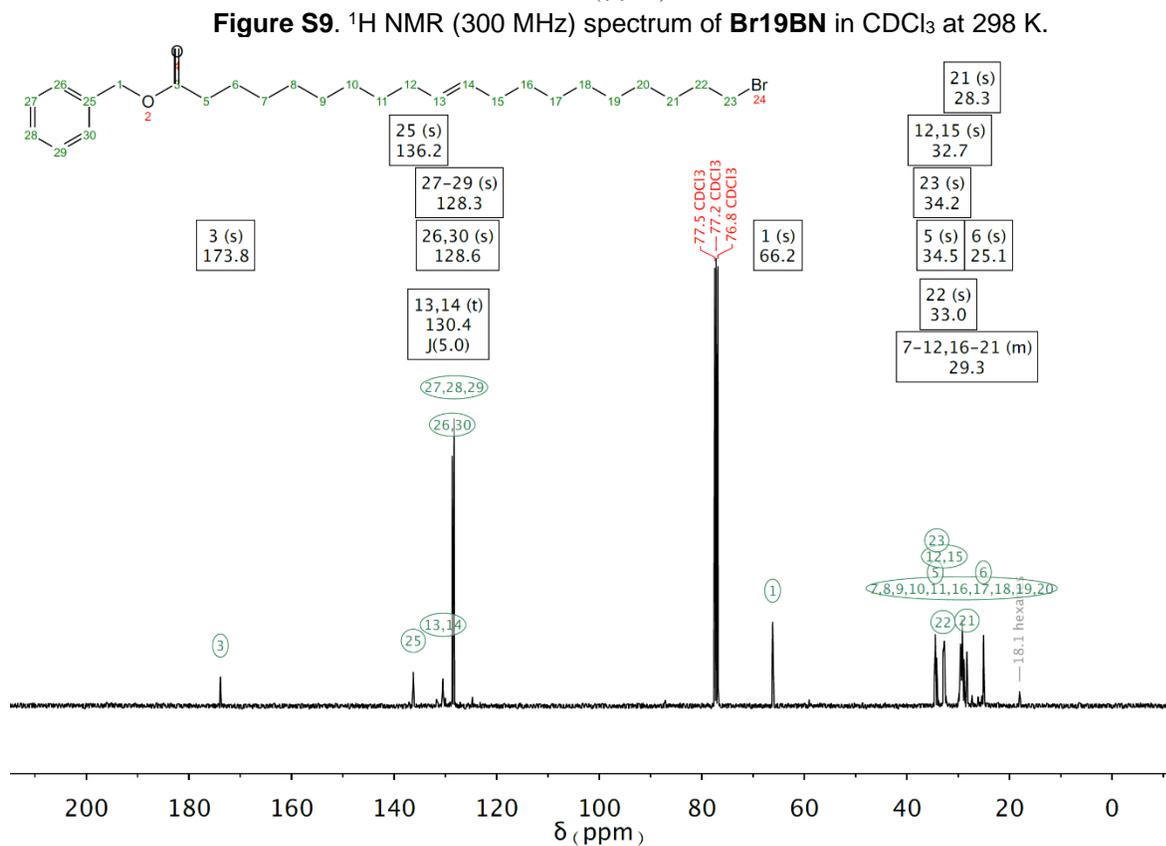

**Figure S10**. ¹³C NMR (101 MHz) spectrum of **Br19BN** in CDCl₃ at 298 K.

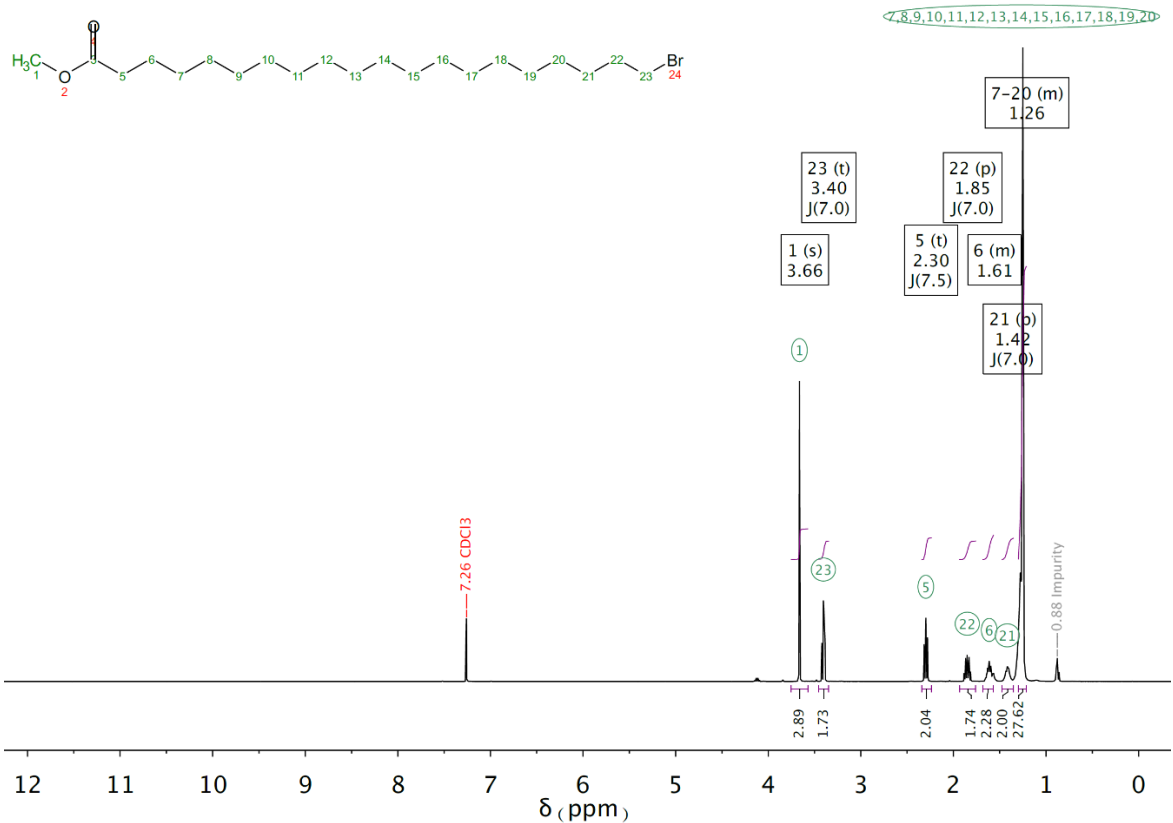

**Figure S11**. ¹H NMR (400 MHz) spectrum of **Br19ME** in CDCl₃ at 298 K.

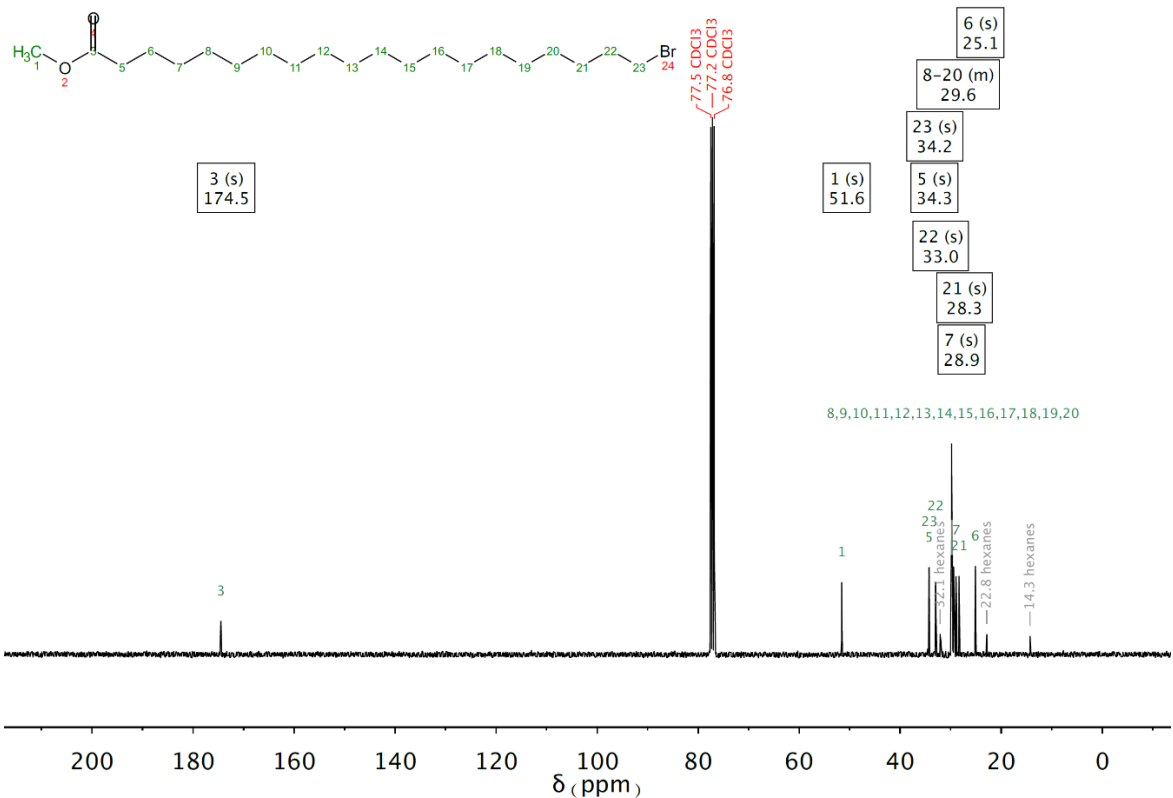

**Figure S12**. ¹³C NMR (101 MHz) spectrum of **Br19ME** in CDCl₃ at 298 K.

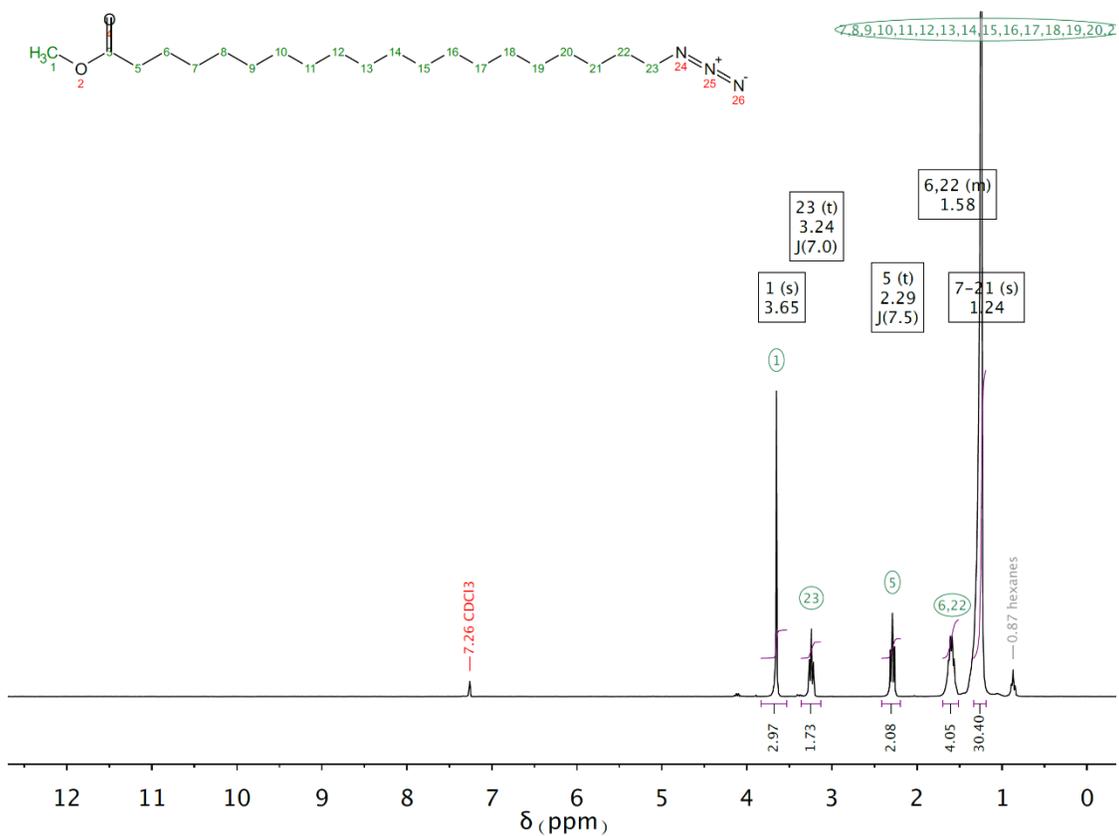

Figure S13. ¹H NMR (300 MHz) spectrum of **az19ME** in CDCl₃ at 298 K.

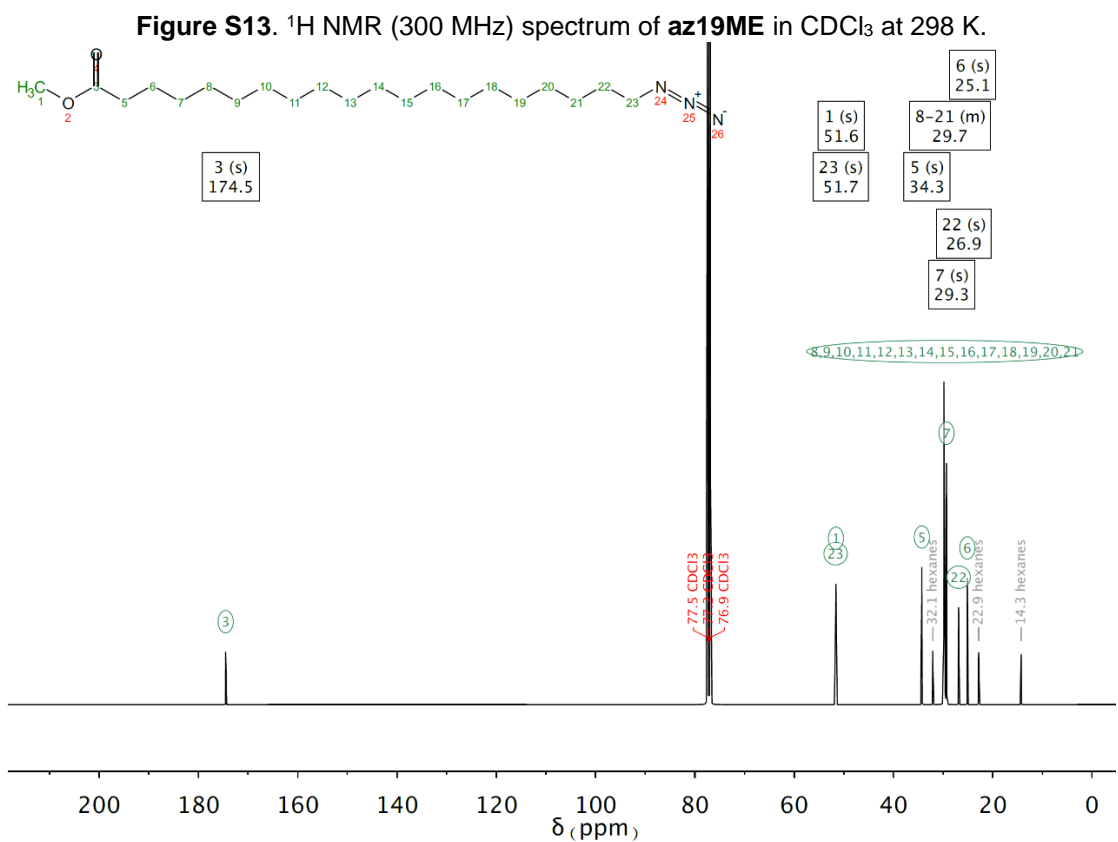

Figure S14. ¹³C NMR (101 MHz) spectrum of **az19ME** in CDCl₃ at 298 K.

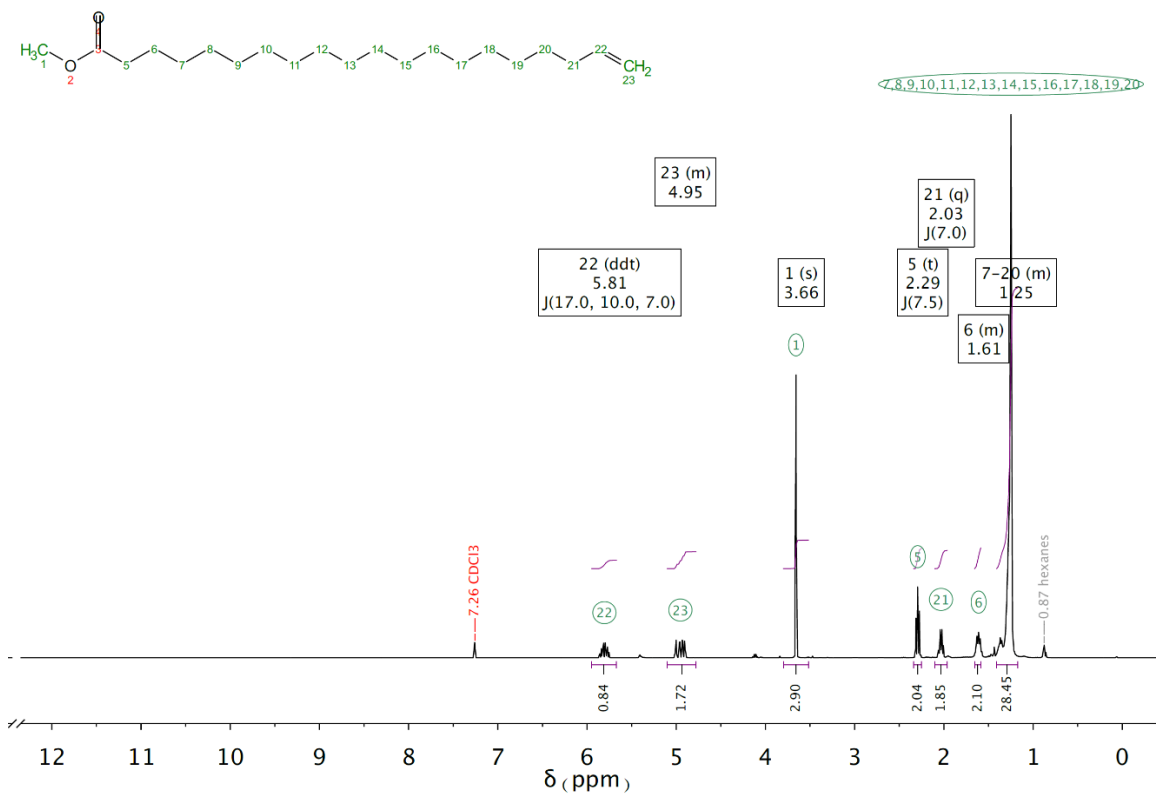

**Figure S15**. ¹H NMR (400 MHz) spectrum of **alkene19ME** in CDCl$_3$ at 298 K.

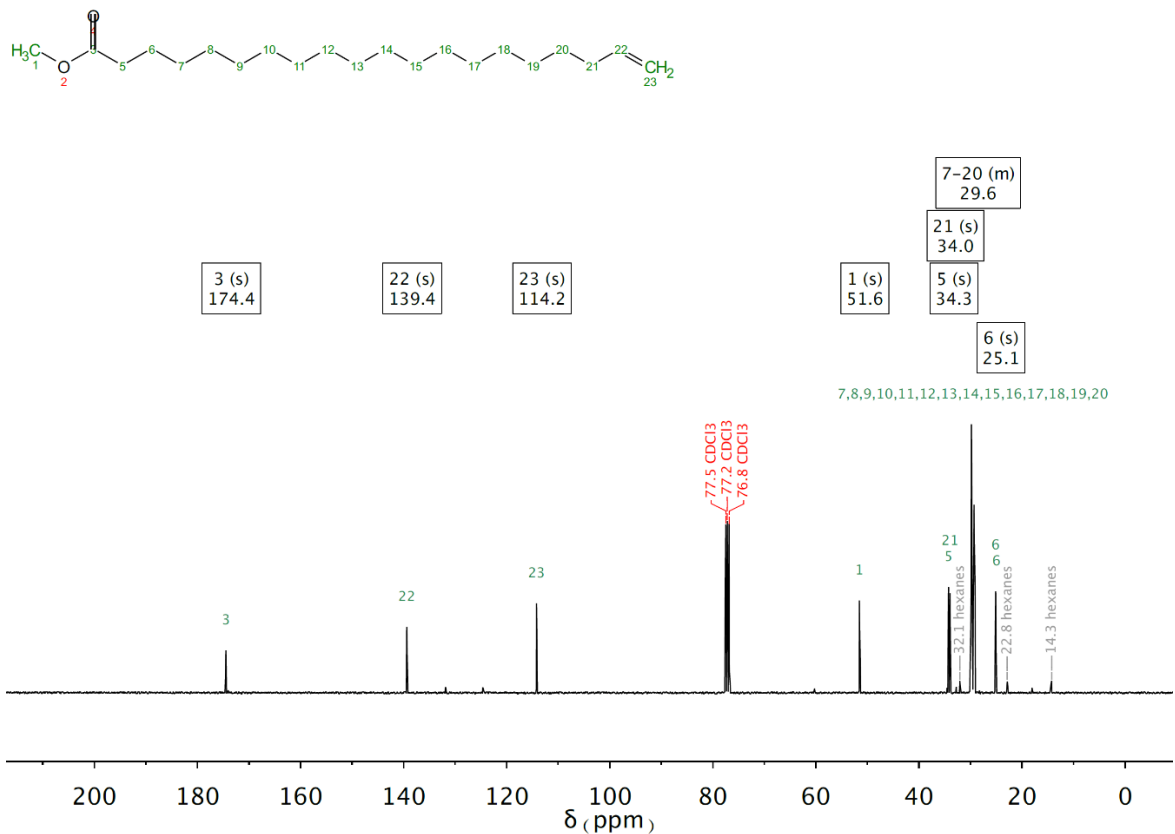

**Figure S16**. ¹³C NMR (101 MHz) spectrum of **alkene19ME** in CDCl$_3$ at 298 K.

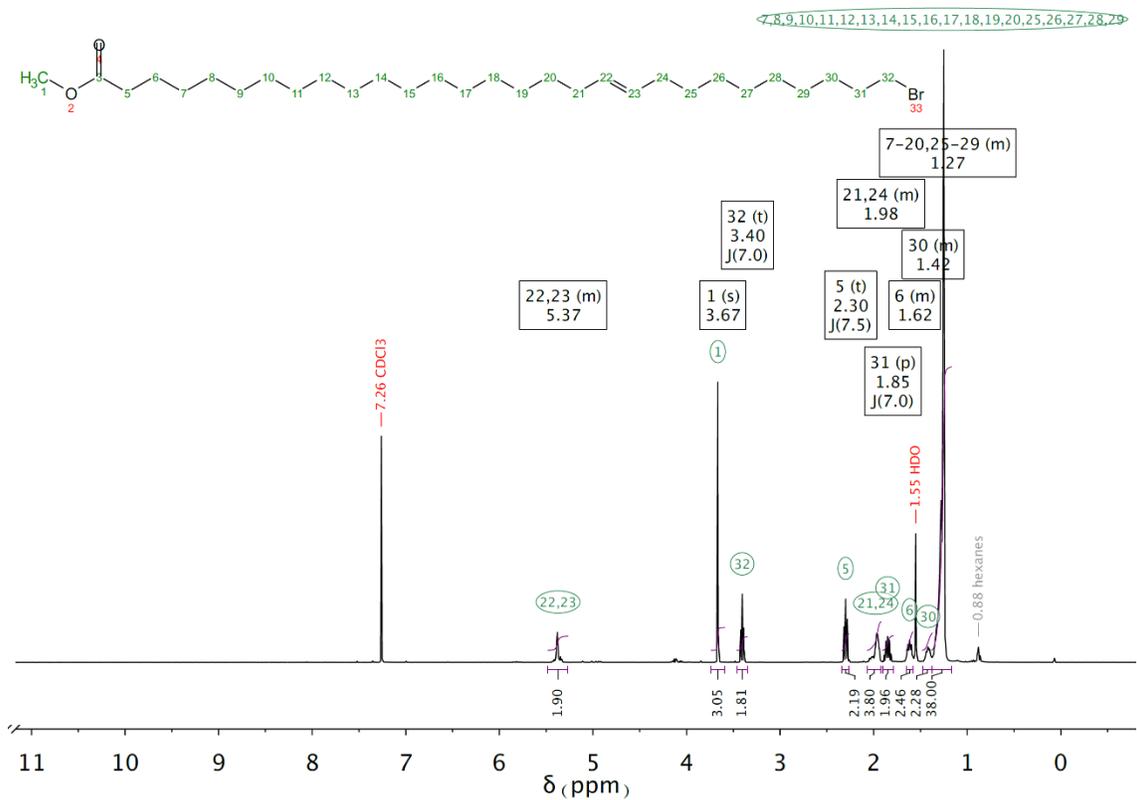

**Figure S17**. ¹H NMR (400 MHz) spectrum of **Br28ME (unsat.)** in CDCl₃ at 298 K.

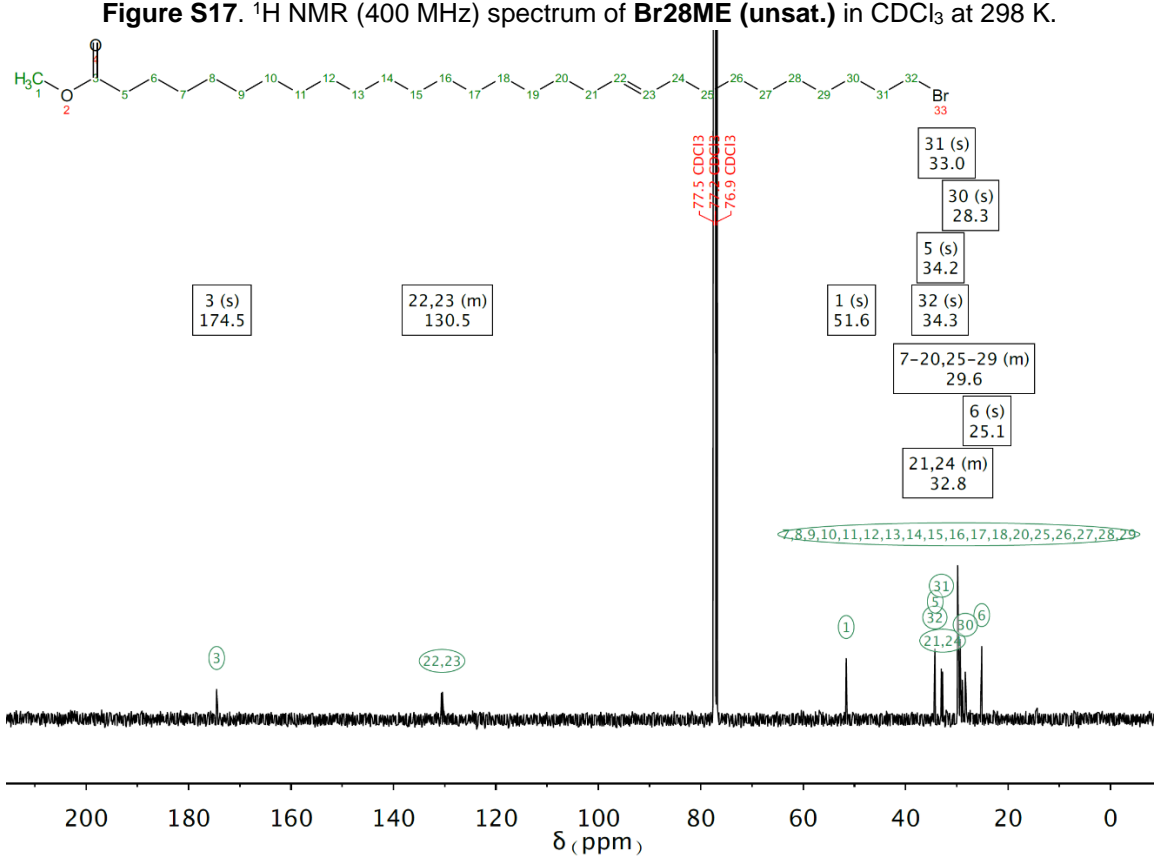

**Figure S18**. ¹³C NMR (101 MHz) spectrum of **Br28ME (unsat.)** in CDCl₃ at 298 K.

**Figure S19.** $^1$H NMR (400 MHz) spectrum of **Br28ME (sat.)** in CDCl$_3$ at 298 K.

**Figure S20.** $^{13}$C NMR (101 MHz) spectrum of **Br28ME (sat.)** in CDCl$_3$ at 298 K.

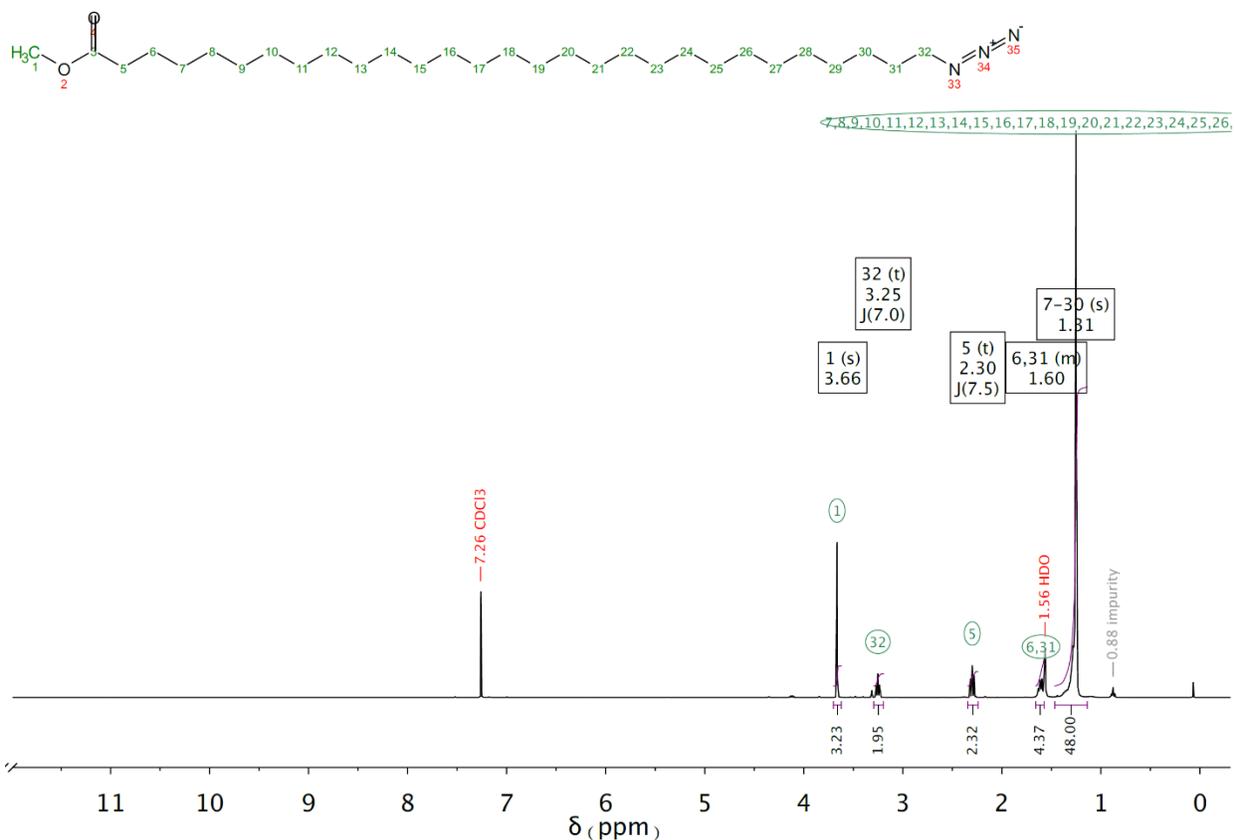

Figure S221. ¹H NMR (400 MHz) spectrum of **az28ME** in CDCl$_3$ at 298 K.

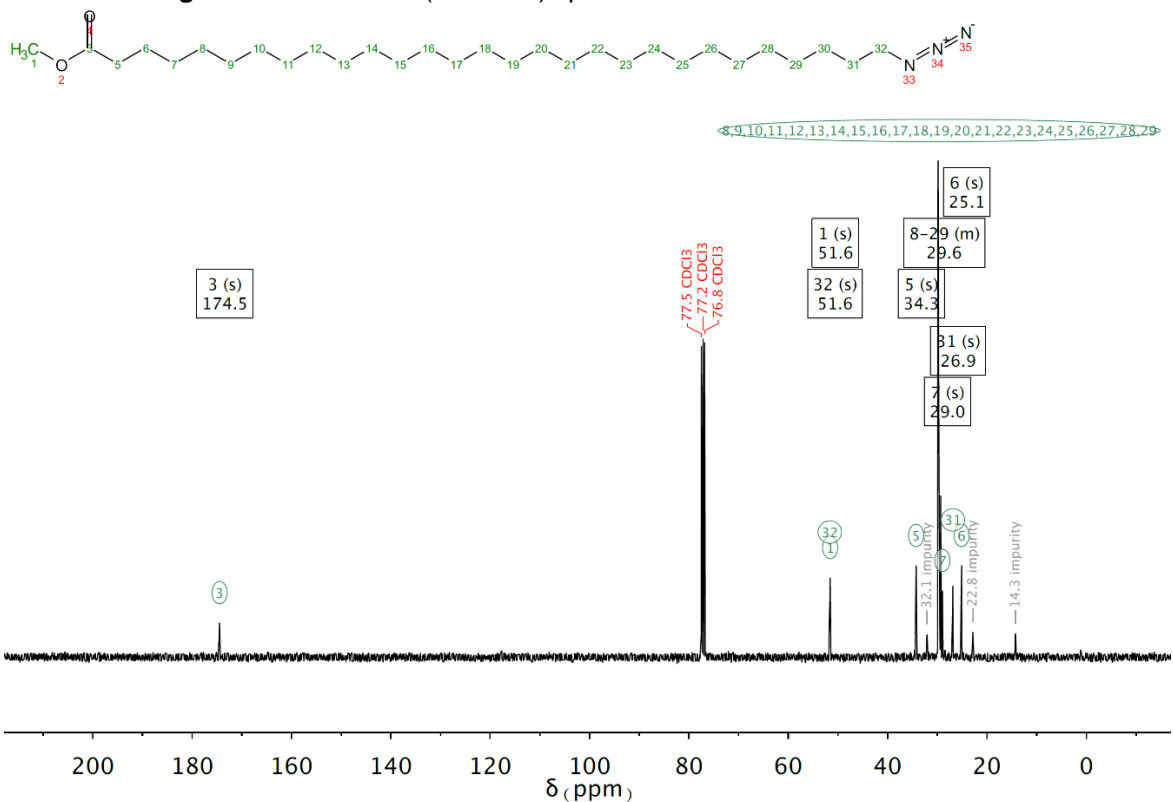

Figure S22. ¹³C NMR (101 MHz) spectrum of **az28ME** in CDCl$_3$ at 298 K.

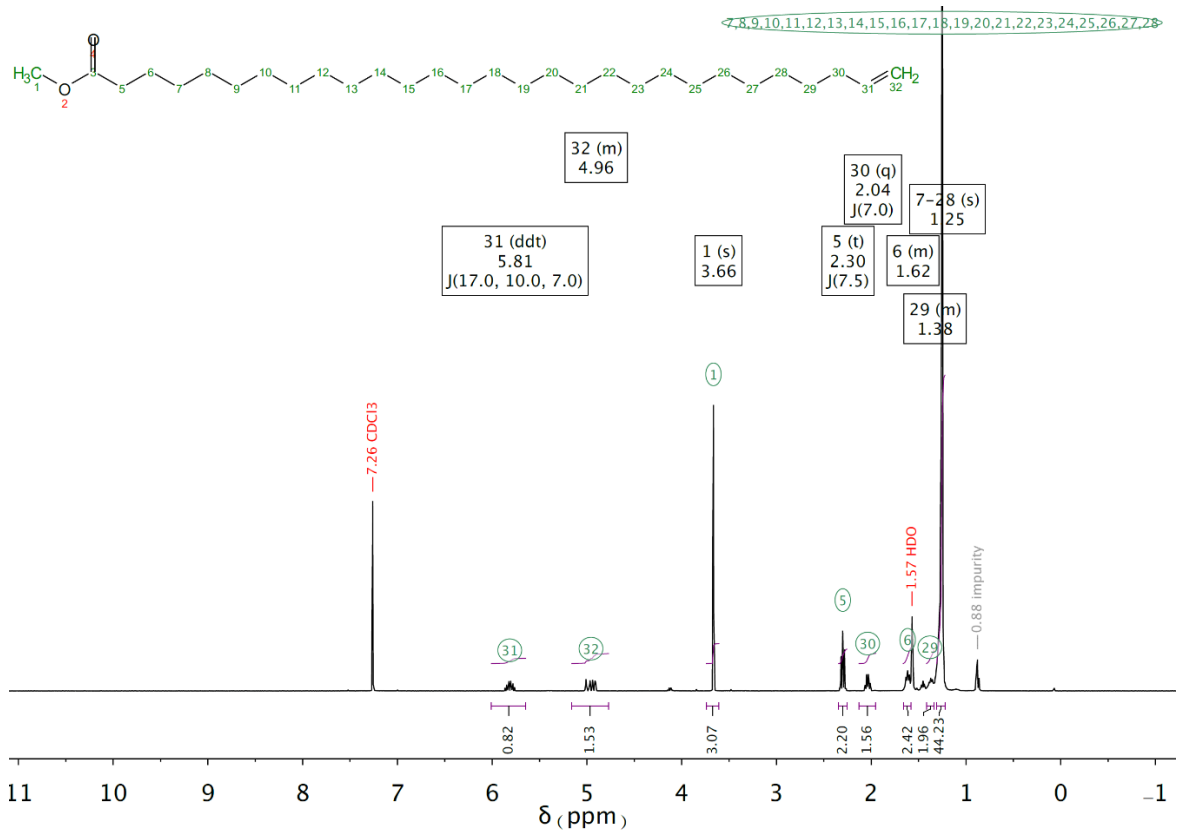

**Figure S222.** $^1$H NMR (400 MHz) spectrum of **alkene28ME** in CDCl$_3$ at 298 K.

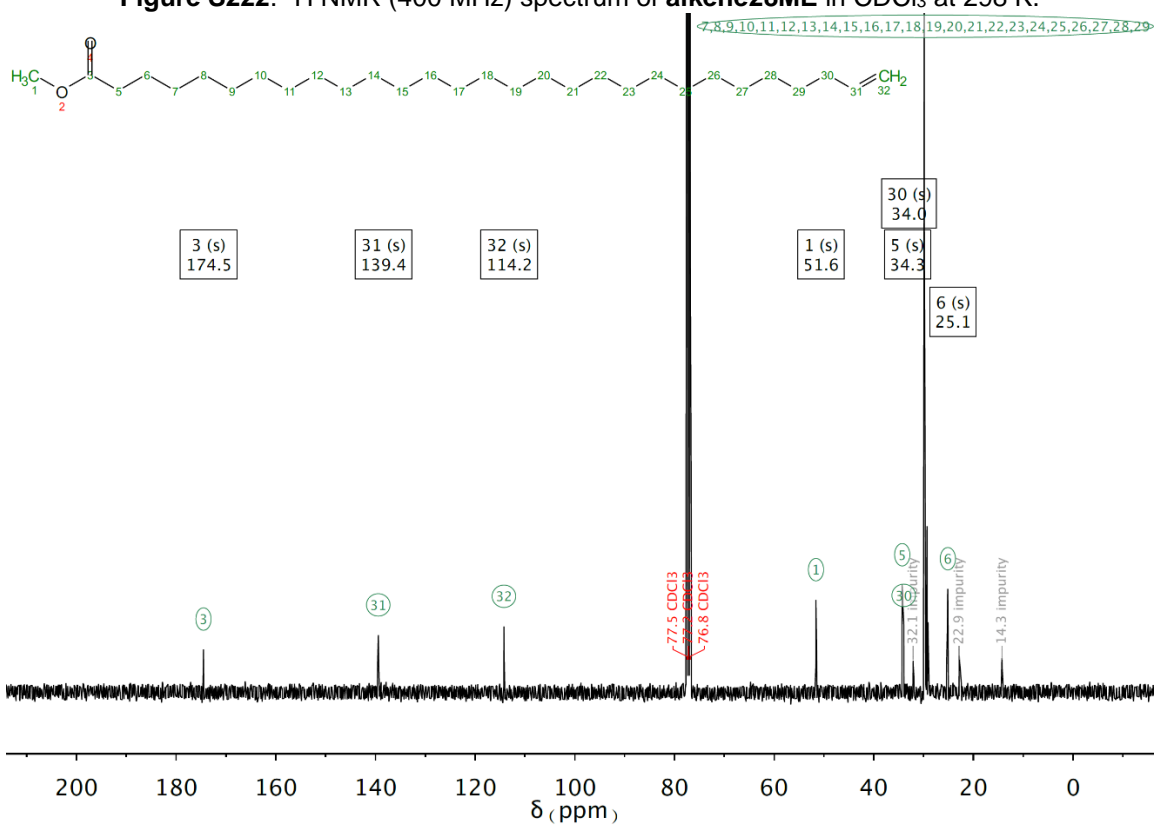

**Figure S24.** $^{13}$C NMR (101 MHz) spectrum of **alkene28ME** in CDCl$_3$ at 298 K.

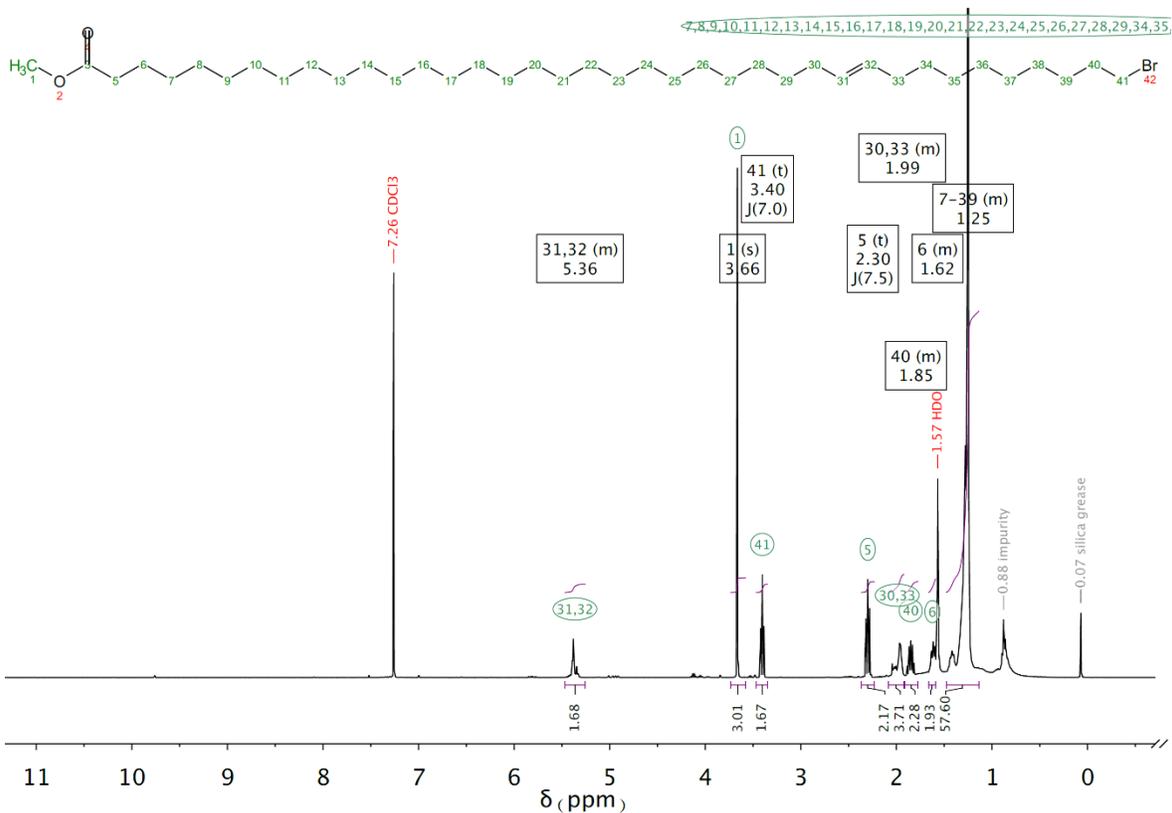

Figure S25. ¹H NMR (400 MHz) spectrum of **Br37ME (unsat.)** in CDCl₃ at 298 K.

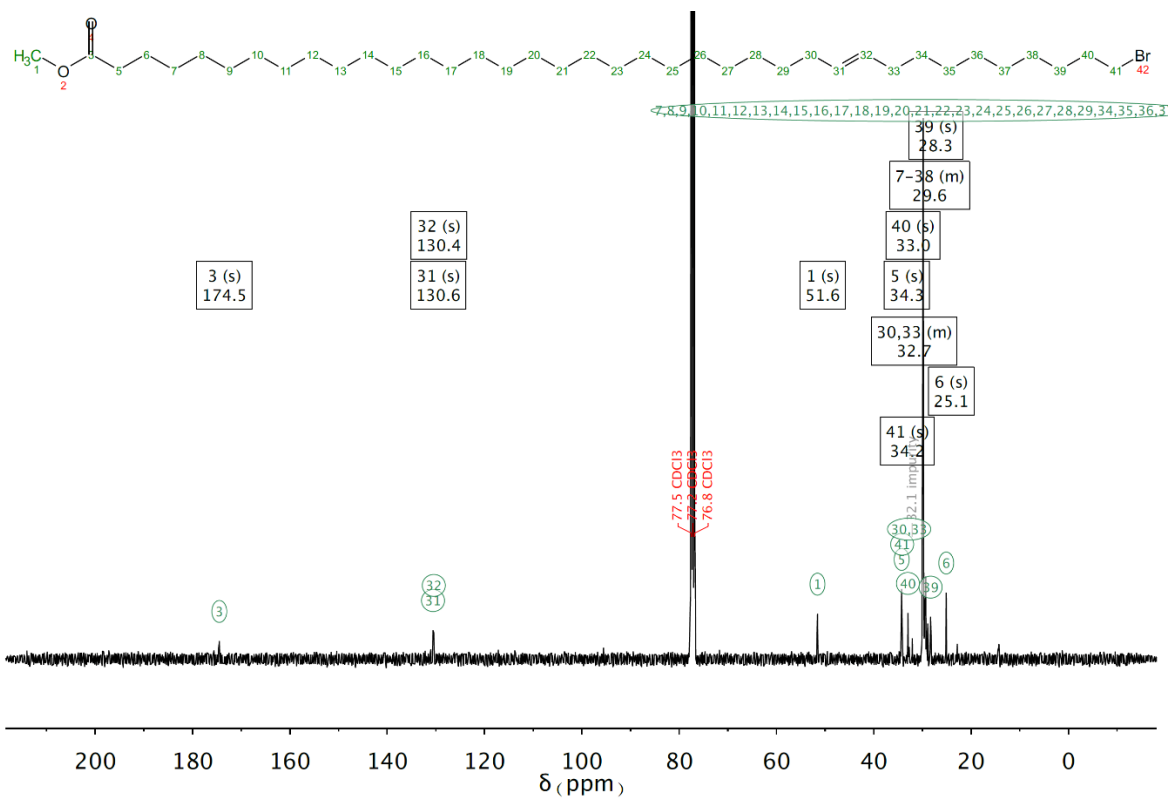

Figure S26. ¹³C NMR (101 MHz) spectrum of **Br37ME (unsat.)** in CDCl₃ at 298 K.

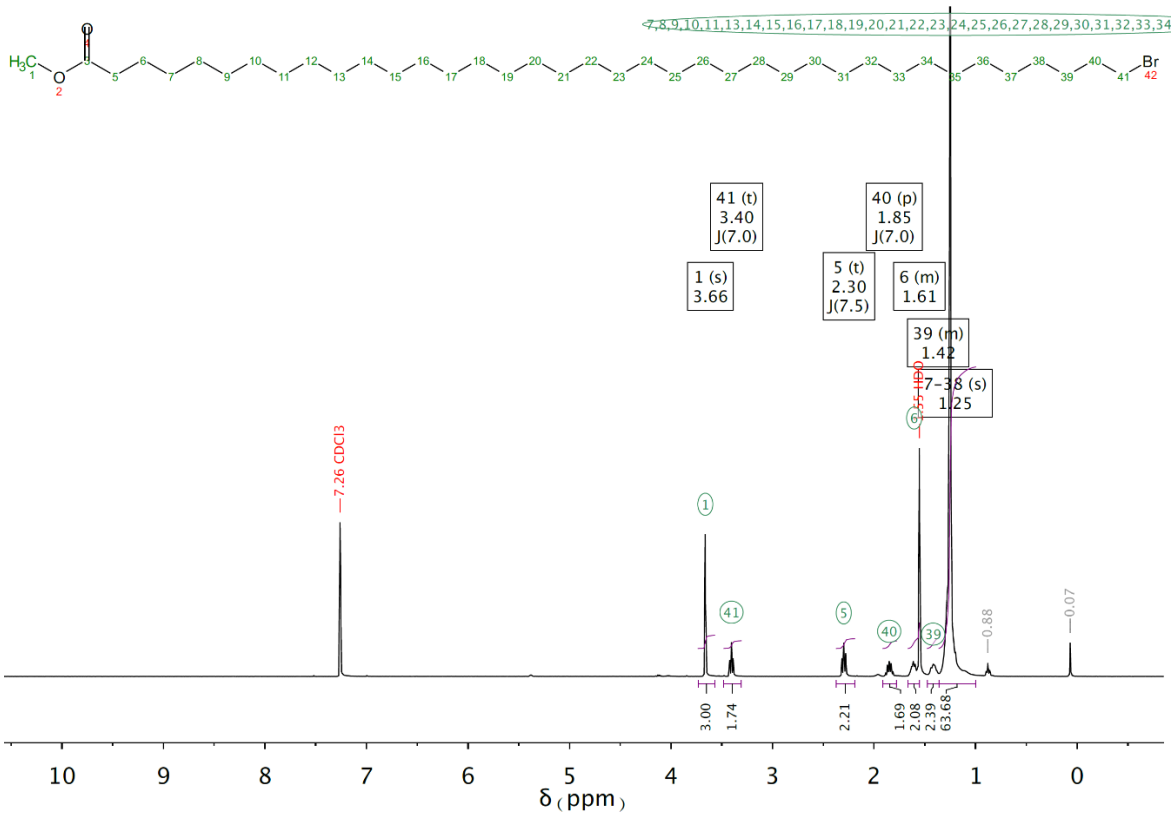

Figure S27. ¹H NMR (400 MHz) spectrum of **Br37ME (sat.)** in CDCl$_3$ at 298 K.

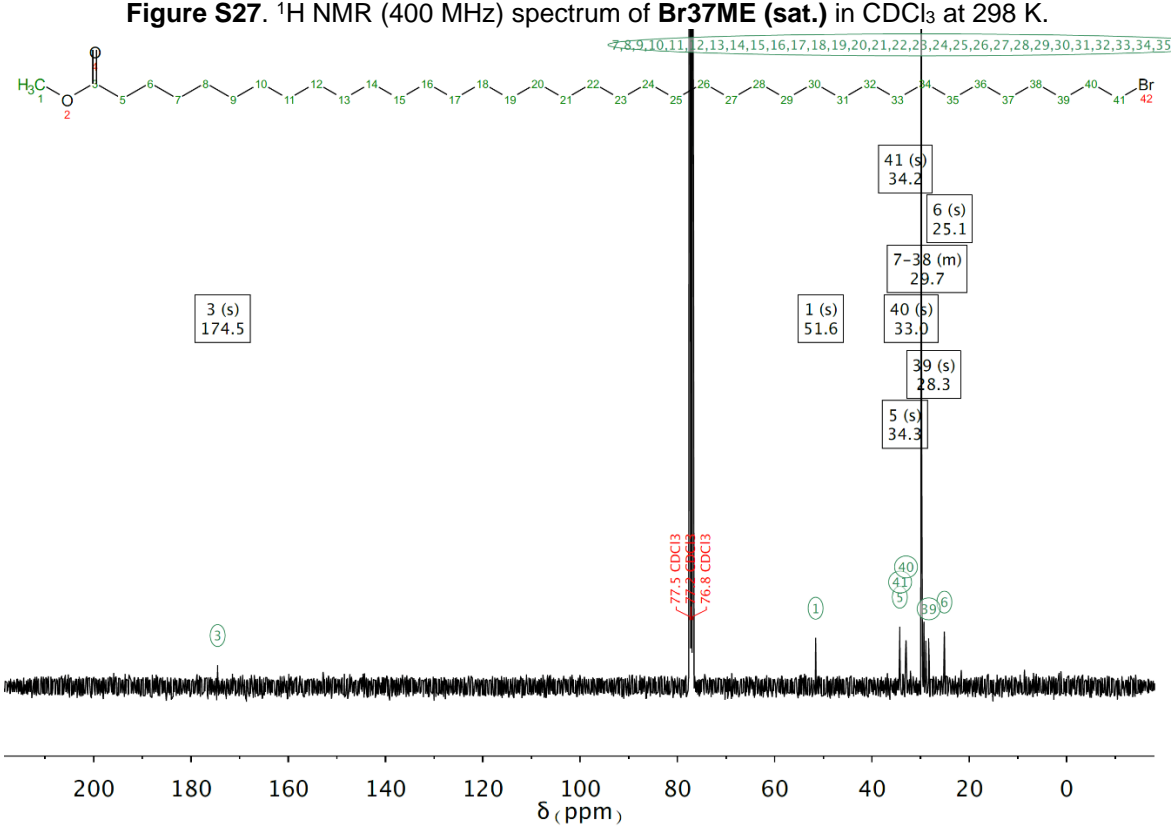

Figure S28. ¹³C NMR (101 MHz) spectrum of **Br37ME (sat.)** in CDCl$_3$ at 298 K.

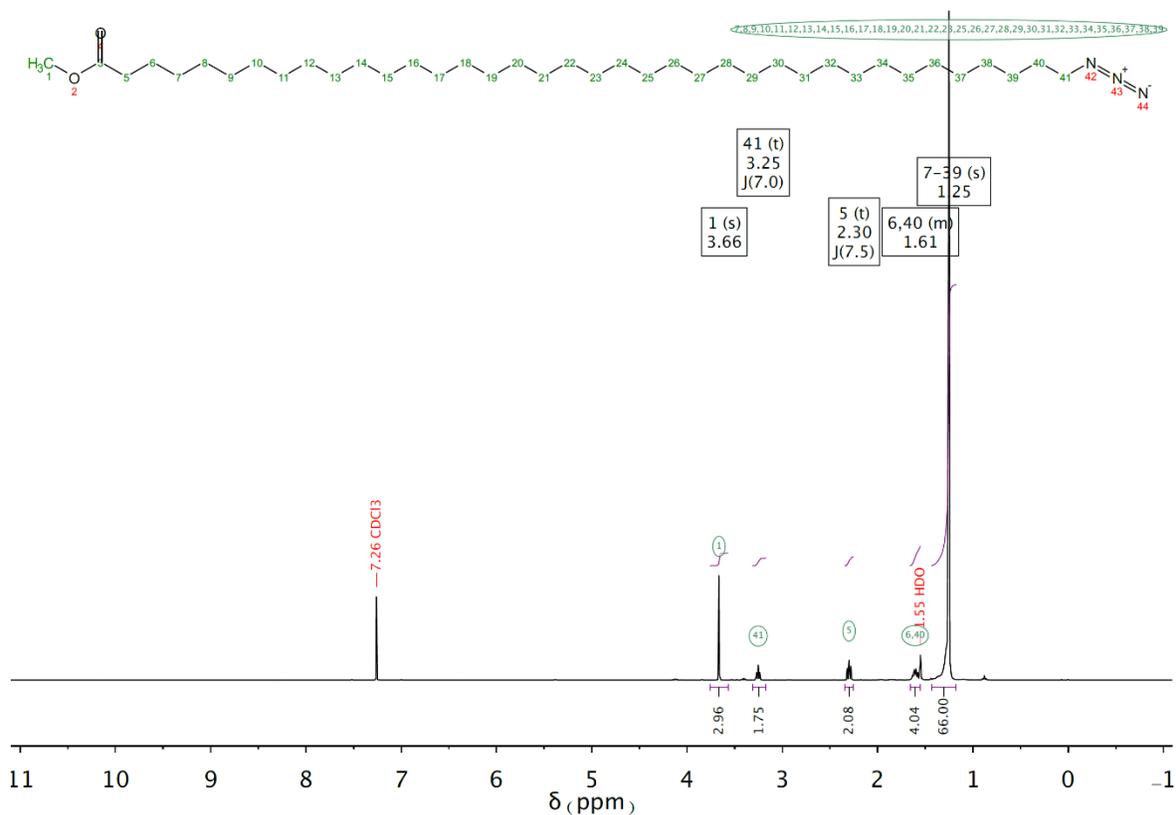

**Figure S29**. $^1$H NMR (400 MHz) spectrum of **az37ME** in CDCl$_3$ at 298 K.

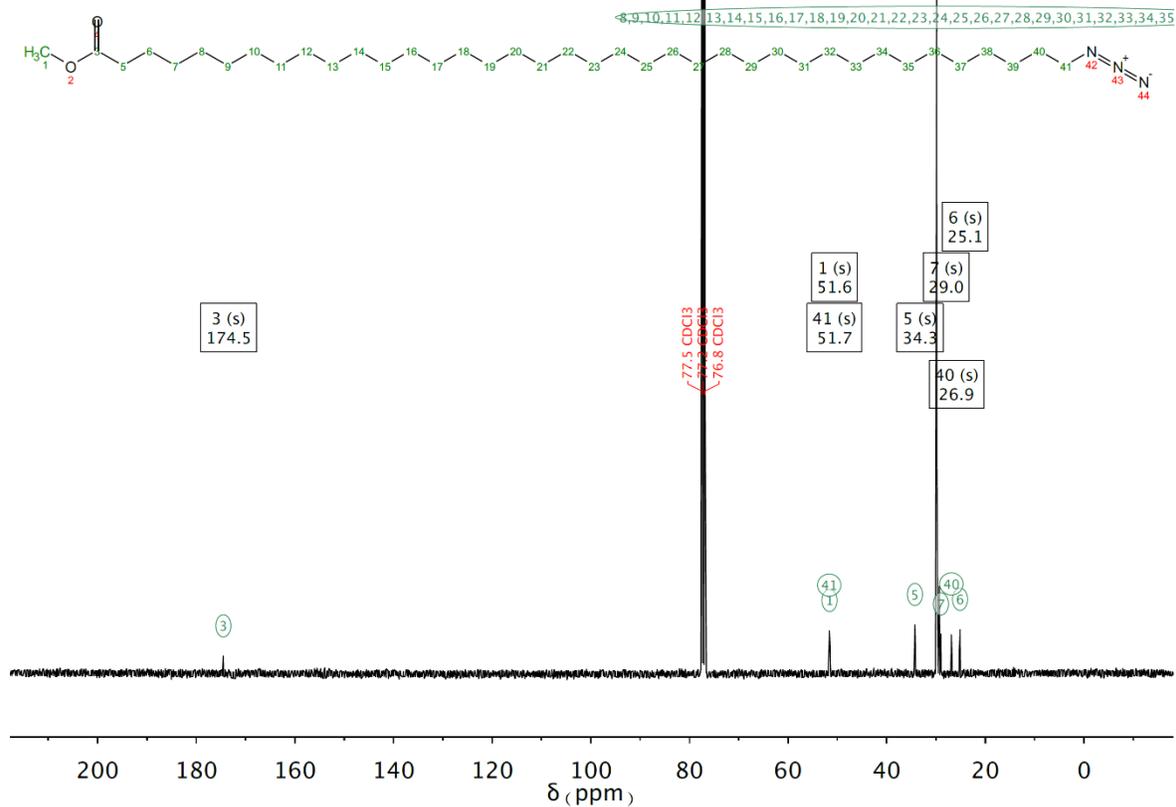

**Figure S230**. $^{13}$C NMR (101 MHz) spectrum of **az37ME** in CDCl$_3$ at 298 K.

## S2. Additional modeling data

Figure S31A shows computed waiting time dependences for the whole range of chain lengths. Thick green lines show the fits with an asymmetric double–sigmoidal function in the vicinity of the maximum. Figure S31B illustrates that the computed $T_{max}$ values are not monotonic with the chain length caused by a small density of states of a chain band that start contributing at specific chain lengths. The W and Ro bands show significant $T_{max}$ fluctuations at $n \sim 12$ (W) and 25 (Ro, Figure S31B). In an attempt to reduce the fluctuations, we introduced inhomogeneity for the tag-group frequency with respect to the chain state site frequency by averaging the dynamics for five tag frequency implementations within inhomogeneity of 12 cm$^{-1}$. Incorporation of the inhomogeneity led to smoother waiting-time dependences, especially for shorter chains, but could not eliminate the "noise" associated with the low chain-state density effect (Fig. S31). Note that larger inhomogeneity values were tested as well but did not reduce the noise in the computed $T_{max}$ values.

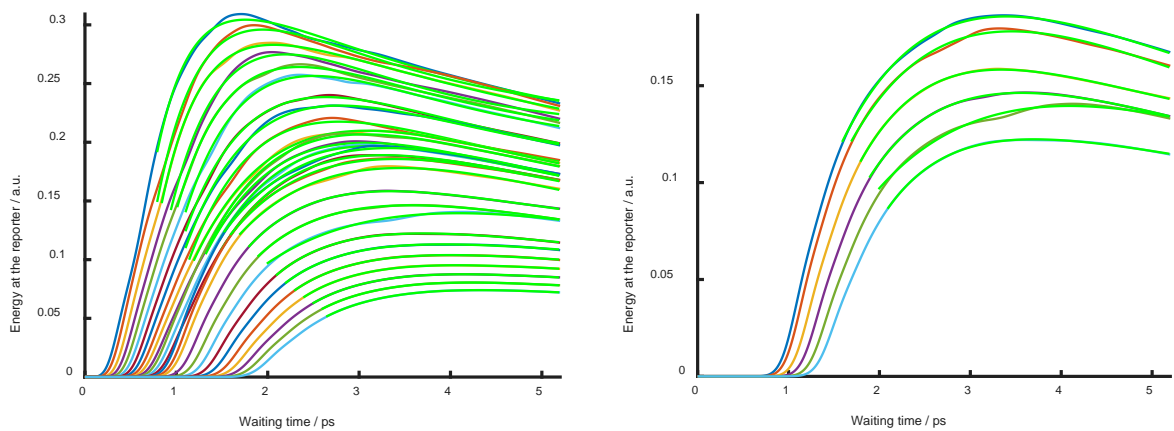

**Figure S31.** Computed waiting time dependences (**A**) for the whole range of chain lengths of $n$ = 3, 4, 5, 6, 7, 8, 9, 10, 11, 12, 13, 14, 15, 16, 17, 18, 19, 21, 23, 25, 27, 29, 31, 33, 35, 37, 39 and (**B**) for six selected lengths: $n$ = 18, 19, 21, 23, 25, and 27. The rising portion of the curves in panel **B** depends monotonically on the chain length. Thick green lines show the fits with an asymmetric double sigmoidal function in the vicinity of the maximum, $y = y_0 + A \left( \dfrac{1}{1+e^{-(x-xc+\frac{w1}{2})/w2}} \right) \left( 1 - \dfrac{1}{1+e^{-(x-xc+\frac{w1}{2})/w3}} \right)$, performed in Origin software.

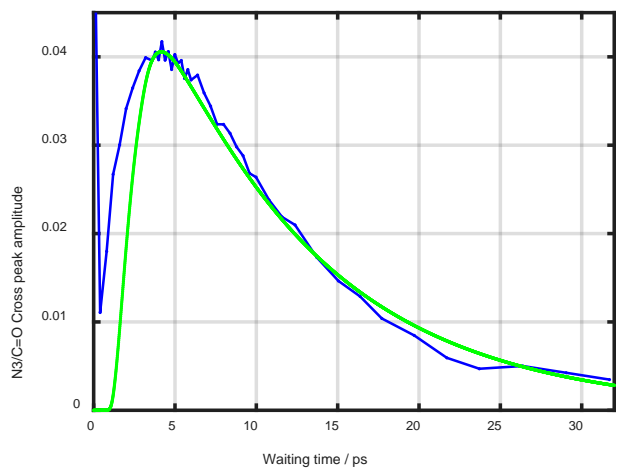

**Figure S32.** Experimental (blue line) and computed (green line) and waiting time dependences for *n* = 19.

**References**

(1) Xie, Y.; Peng, C.; Gao, Y.; Liu, X.; Liu, T.; Joy, A. Mannose-based graft polyesters with tunable binding affinity to concanavalin A. *Journal of Polymer Science Part A: Polymer Chemistry* **2017,** *55* (23), 3908.
(2) Gonzalez, C.; Sanchez, A.; Collins, J.; Lisova, K.; Lee, J. T.; Michael van Dam, R.; Alejandro Barbieri, M.; Ramachandran, C.; Wnuk, S. F. The 4-N-acyl and 4-N-alkyl gemcitabine analogues with silicon-fluoride-acceptor: Application to 18F-Radiolabeling. *European Journal of Medicinal Chemistry* **2018,** *148*, 314.
(3) Hang, H. C.; Geutjes, E.-J.; Grotenbreg, G.; Pollington, A. M.; Bijlmakers, M. J.; Ploegh, H. L. Chemical Probes for the Rapid Detection of Fatty-Acylated Proteins in Mammalian Cells. *Journal of the American Chemical Society* **2007,** *129* (10), 2744.